\documentclass[preprint2]{emulateapj}
\begin{document}

\title{Morphology Dependence Of Stellar Age in Quenched Galaxies at Redshift $\sim
  1.2$:  Massive Compact Galaxies Are Older Than More Extended Ones. } 

\author{Christina C. Williams\altaffilmark{1}, 
Mauro Giavalisco\altaffilmark{2}, 
Rachel Bezanson\altaffilmark{1,3},
Nico Cappelluti\altaffilmark{4,5},
Paolo Cassata\altaffilmark{6},
Teng Liu\altaffilmark{7,2},
Bomee Lee\altaffilmark{2},
Elena Tundo\altaffilmark{8},
Eros Vanzella\altaffilmark{9}
}

\altaffiltext{1}{Steward Observatory, 933 N. Cherry Ave., University of Arizona, Tucson, AZ 85721, USA,
ccwilliams@email.arizona.edu}
\altaffiltext{2}{Department of Astronomy, University of Massachusetts, 710 North Pleasant Street, Amherst, MA 01003, USA}
\altaffiltext{3}{Hubble Fellow}
\altaffiltext{4}{Department of Physics, Yale University, P.O. Box 208121, New Haven, CT 06520, USA}
\altaffiltext{5}{Yale Center for Astronomy \& Astrophysics, Physics Department, P.O. Box 208120, New Haven, CT 06520, USA}
\altaffiltext{6}{Instituto de F\'isica y Astronom\'ia, Facultad de Ciencias, Universidad de Valpara\'iso, Gran Breta\~na 1111, Valpara\'iso, Chile} 
\altaffiltext{7}{CAS Key Laboratory for Research in Galaxies and Cosmology, Department of Astronomy, University of Science and Technology of China, 230026 Hefei, Anhui, China}
\altaffiltext{8}{Visiting scholar, Department of Astronomy, University of Massachusetts, 710 North Pleasant Street, Amherst, MA 01003, USA}
\altaffiltext{9}{INAF - Osservatorio Astronomico di Bologna, Bologna, Italy}

\begin{abstract}
We report the detection of morphology--dependent stellar age in massive quenched
 galaxies (QGs) at z$\sim$1.2. The sense of the dependence is
that compact QGs are 0.5--2 Gyr older than normal-sized ones. The
evidence comes from three different age indicators, D$_n4000$, H$_{\delta}$
and fits to spectral synthesis models, applied to their stacked optical spectra. All age indicators consistently show that the stellar populations of compact QGs are older than their normally--sized counterparts. We detect weak [\ion{O}{2}] emission in a 
 fraction of QGs, and the strength of the
line, when present, is similar between the two samples; however, compact
galaxies exhibit significantly lower frequency of [\ion{O}{2}] emission than
normal ones. A fraction of both samples are individually detected in 7--Ms Chandra X--ray images (luminosities$\sim10^{40}$--$10^{41}$
erg/sec). 7--Ms stacks of non-detected galaxies show similarly low luminosities in the soft band only, consistent with a hot gas origin for the X--ray emission.
 While both [\ion{O}{2}] emitters and non-emitters are also X--ray sources
among normal galaxies, no compact galaxy with [\ion{O}{2}] emission is an
X--ray source, arguing against an AGN powering the line in compact galaxies. 
We interpret the [\ion{O}{2}] properties as further evidence that compact galaxies are older and further along into the process of quenching star--formation and suppressing gas accretion.  Finally, we argue that the older age of compact QGs is evidence
of progenitor bias: compact QGs simply reflect the smaller sizes of galaxies at their earlier quenching epoch, with stellar density
most likely having nothing directly to do with cessation of star--formation.
\end{abstract}

\section{Introduction}
\label{Introduction}

The formation and evolution of massive early-type galaxies remains poorly understood, despite much recent progress.
Constraints from the
local Universe indicate that their stellar ages are very old ($>$10 Gyr),
indicating that they formed the bulk of their stellar masses at z$>$2
\citep{Bower1992, Renzini1993, vanDokkumEllis2003, Heavens2004, Renzini2006},
subsequently quenching star-formation, and remaining quenched, until the
present. Constraints on stellar abundance ratios (high $\alpha$/Fe) 
indicate  their star-formation took place on short timescales
\citep{Thomas2005, Thomas2010, Renzini2006}. Additionally, it has been
observed that galaxy morphology and star-formation properties are highly
correlated, such that this quenched nature in massive galaxies appears
coincident with morphological transformation to ellipsoidal 
 stellar structure \citep[e.g.][]{Strateva2001,Kauffmann2003b,Franx2008}. Despite efforts to
study this transition from star-forming galaxy to quenched ellipsoid, we have
gained very little insight into both the transformational quenching process,
as well as the mechanisms preventing further star-formation for the majority
of the Universe's history.

Of particular importance to this effort are constraints from observing the
progenitors of these massive early-type galaxies at z$>$1, shortly after their
transformation from star-forming galaxy. Recently quenched galaxies (QGs) begin to
appear in large numbers at z$\sim$2 \citep{Cimatti2008, vanDokkum2008, Cassata2011,Cassata2013}, and have
been observed even out to z$\sim$3-4 \citep[][]{Fontana2009, Guo2012, Gobat2012, Muzzin2013, Stefanon2013,
  Straatman2014}. The properties of these quenched high-redshift galaxies
provide significant insight into both the formation process of the galaxies
during the star-formation phase, as well as the quenching mechanisms causing
their transformation. The most striking feature of these recently quenched
galaxies at high-redshift is their stellar structure; while already having
built up a similar amount of stellar mass as their z$\sim$0 counterparts, they
are remarkably compact in stellar density \citep{Daddi2005, Trujillo2006, Bundy2006, Zirm2007,
  Toft2007, vanDokkum2008, Cimatti2008, vanderwel2008, Bezanson2009, Saracco2009, Damjanov2009, Williams2010}. The overwhelming majority of QGs
($>$80\%) at z$>$1.5 have stellar densities higher than the lower 1$\sigma$ of
passive (early-type) galaxies at z$\sim$0 at the same stellar mass \citep{Cassata2013}. 
Additionally, they are much smaller than the
majority of massive star-forming galaxies at the same epoch \citep{vanderWel2014}, and in
fact, one of the strongest predictors of quiescence among high-redshift
galaxies is centrally concentrated light \citep[i.e. a measure of
  compactness;][]{Franx2008, Bell2012, Omand2014, Teimoorinia2016, Whitaker2016}. It appears, therefore, that
compactness and quenched nature at high-redshift are inextricably linked.

However, the physical reason for this correlation is also poorly understood. Does the existence of
the compact QGs at high-redshift imply something very fundamental about
quenching? In other words, does some physical process associated with stellar
compactness predispose galaxies to quench? Alternatively, are the earliest galaxies to
form in the Universe and complete their evolution simply the densest because
the Universe was denser at early times \citep[e.g.][]{LillyCarollo2016}), or, because of some highly dissipative gaseous process that could take place predominantly at high redshift \citep[e.g.][]{Dekel2009, Johansson2012,DekelBurkert2014,Zolotov2015, Ceverino2015}.

There are some physically motivated reasons to
believe that the former may be true. First, high stellar density implies a
previous epoch of high surface density of star-formation, which would mean a
higher energy input into the interstellar medium (ISM) of compact galaxies,
than might be present in larger, extended galaxies. \citet{Hopkins2010} have
made this argument, based on the observation that there appears to be a
maximum stellar density of {\it any} structure in the Universe ($\Sigma \sim$ 10$^{11}$ M$_{\odot}$ kpc$^{-2}$). This limit in stellar density exists despite covering
 8 orders of magnitude in stellar mass, from star
clusters within galaxies, to the z$>$2 compact QGs. This empirical limit argues
for some stellar feedback process, such as massive stellar winds, that
truncate the star-formation and prevent further growth beyond this density 
limit. Studies of objects with high surface density of star-formation, where
such extreme stellar feedback might be expected, have in fact found evidence
of feedback in the form of very fast ($\sim$1000 km/s) galactic-scale outflows
from extremely compact star-forming regions that approach the Eddington limit
 \citep{DiamondStanic2012, Sell2014}. 

The second, alternative scenario exists, that this connection is a very simple
 consequence of the size-evolution of star-forming galaxies, whose radii (at fixed mass) are observed to decrease with increasing redshift \citep[e.g.][]{vanderWel2014}.
The most massive galaxies in the early Universe have formed the
earliest in cosmic time, and therefore evolve to the end of their star-forming
lifetime earliest. In this scenario, the density of the parent halo of the quenched population at any redshift
reflects the density of the Universe at its formation epoch
\cite[e.g.][]{Mo1998}, and therefore will progressively increase in size (and
therefore stellar density) over time.  Such a scenario, known as progenitor
bias \citep[as described by][]{LillyCarollo2016}, would also contribute to the increasing size evolution of QGs over cosmic time
\citep{Valentinuzzi2010a, Valentinuzzi2010b, Poggianti2013, Carollo2013, LillyCarollo2016} \citep[see also][]{Bezanson2009}. In this scenario, the significance of
compactness is irrelevant for the quenching, and rather, galaxies quench when
they've reached sufficient mass that they no longer support
star-formation. The quenching mechanism then may be related to halo mass, or
some other mass related mechanism to cut of gas supply for future
star-formation \citep[e.g.][]{Peng2010, BirnboimDekel2003, DekelBirnboim2006}.
 Distinguishing between these scenarios is highly
important to understanding the evolution of early-type galaxies. 

Each of these two scenarios have empirical predictions for the properties of QGs.  In the stellar-density regulated star-formation scenario, galaxies with high enough surface density of star-formation will quench, and produce remnants with high stellar density. Thus at any given epoch, the most recently quenched objects should also be the densest \citep[e.g.][]{Whitaker2012}. There is no explicit prediction for a trend of stellar density with stellar age or mass \citep[e.g.][]{Hopkins2010}. However, progenitor bias explicitly predicts that stellar age, and stellar density be correlated, with the densest objects also being the oldest at any given epoch \citep[in the absence of size growth via merging;][]{LillyCarollo2016}.
In this paper, we seek to distinguish 
 between these
two scenarios, and in the process, gain insight into why galaxies quench their star-formation early in cosmic
time. In Section 2 we present the data used in this study. in Section 3 we
present our results, and in Section 4 we discuss these results in the context
of quenching and the formation of the QGs. Throughout this paper we assume a
cosmology with $\Omega_{\Lambda}$=0.7, $\Omega_{M}$=0.3, and H$_{o}$ = 70
km/s/Mpc.

\begin{figure*} [!t]
\begin{center}
\includegraphics[scale=0.48]{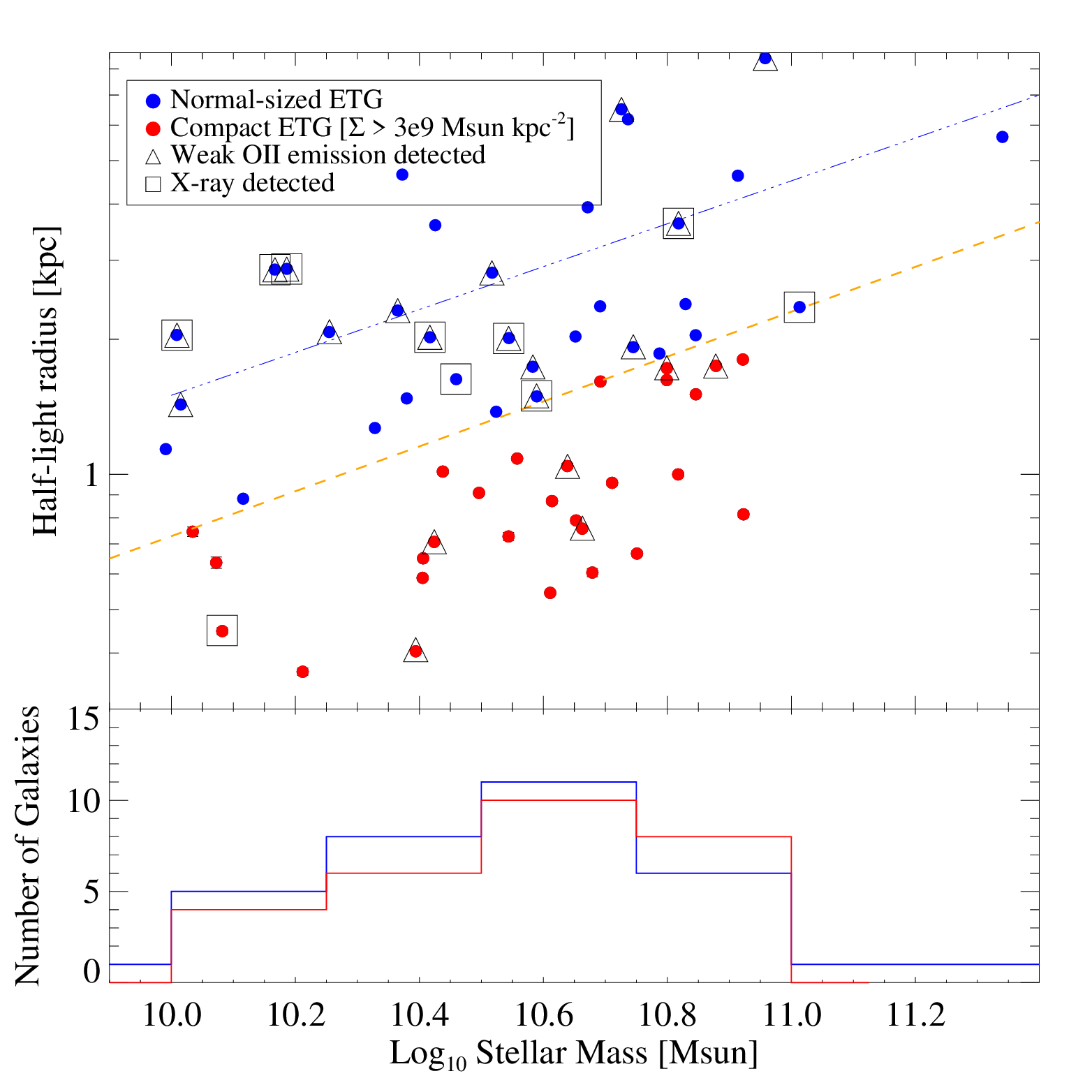}

\caption{The mass-size relation of z$\sim$1.2 QGs in this study. In red are QG defined as compact according to \citet{Cassata2013}, as being below the lower 1-$\sigma$ of the z$\sim$0 early-type galaxy mass-size relation \citep[orange dashed line;][]{Shen2003}. The z$\sim$0 mean early-type galaxy mass-size relation is the blue dot-dashed line \citep{Shen2003}. Blue galaxies are considered normally-sized QG (relative to early-type galaxies at z$\sim$0). Galaxies with triangles designate those in which [\ion{O}{2}]$\lambda$3727 emission was detected. Squares designate those galaxies with X-ray detections. Bottom panel: the mass distributions of each sample are roughly equivalent.} 
\label{masssize}
\end{center}
\end{figure*}

\section{Data}
\label{Data}

\subsection{Samples}
We select QGs from the Cosmic Assembly Near-infrared Deep Extragalactic Legacy Survey (CANDELS) data \citep{Koekemoer2011, Grogin2011} in GOODS-South according to the selection outlined in \citet{Cassata2011}, which identified 179 QGs at z$>$1 with M*$>$10$^{10}$M$_{\odot}$, specific SFR $<$ 10$^{-2}$ Gyr$^{-1}$ and centrally concentrated, spheroidal morphologies. For this study, we have measured the properties of this sample using the CANDELS multi-wavelength photometry \citep{Guo2013} following the SED fitting procedure outlined in B. Lee et al. (in preparation) using the SpeedyMC Bayesian SED-fitting software \citep{Acquaviva2011} where star-formation history is left as a free parameter. 
 We use the morphologies measured from the HST/WFC3 F160W {\it H}-band imaging from CANDELS using GALFIT \citep{Peng2002} presented in \citet{vanderWel2012}. In particular, we define the size to be the circularized half-light radius in kpc,  R$_{eff} = r_{e} \sqrt{b/a}$, where $r_{e}$ is the length of the semi-major axis in arc seconds converted to kpc using the spectroscopic redshift, and $b/a$ is the axis ratio. 

From this parent sample, we identify 61 QGs with complementary publicly available spectra (see next section). We define a 'compactness' cut on the QG sample according to the local z$\sim$0 mass-size relation for local galaxies from the Sloan Digital Sky Survey (SDSS). We define QGs as compact if their size (at a given mass) is smaller than the lower 1-$\sigma$ of the local early-type galaxy size-mass relation \citep{Shen2003}. This roughly corresponds to a stellar mass surface density of $\Sigma \sim$3x10$^{9}$M$_{\odot}$kpc$^{-2}$. We refer to any QG with a stellar density higher than this threshold as compact. QGs that are more extended, and thus similar to the majority of local SDSS early-type galaxies in mass and size, we refer to as 'normal'. The position of the QGs in the mass-size diagram, illustrating the compactness cut from SDSS, are presented in Figure \ref{masssize}. 
There are 28 compact galaxies and 33 normal-sized galaxies. Average redshift for the samples are $<$z$>=$1.22 and 1.13 for compact and normal, respectively.
\subsection{Spectroscopy}
Our spectroscopic data for this sample was obtained at the European Southern Observatory's Very Large Telescope (VLT) as part of the Great Observatories Origins Deep Survey \citep[GOODS;][]{Giavalisco2004} spectroscopic program. In particular, we use spectra from programs with 
FORS2 \citep{Vanzella2005, Vanzella2006, Vanzella2008, Kurk2009, Kurk2013}, and VIMOS \citep{Popesso2009, Balestra2010}. The FORS2 spectra have an instrument resolution of R$\sim$660, which is 13\AA\ at 8600\AA\ observed (average 5.9 at 3900\AA\ restframe; all spectra between redshifts 1 and 1.4).
The spectra we use from VIMOS in the \citet{Popesso2009, Balestra2010} release are all using the medium resolution grism, which has comparable resolution. 
\subsection{Stacking procedure}
\begin{figure*} [!t]
\begin{center}
\includegraphics[scale=0.48, trim=0 250 0 0,clip]{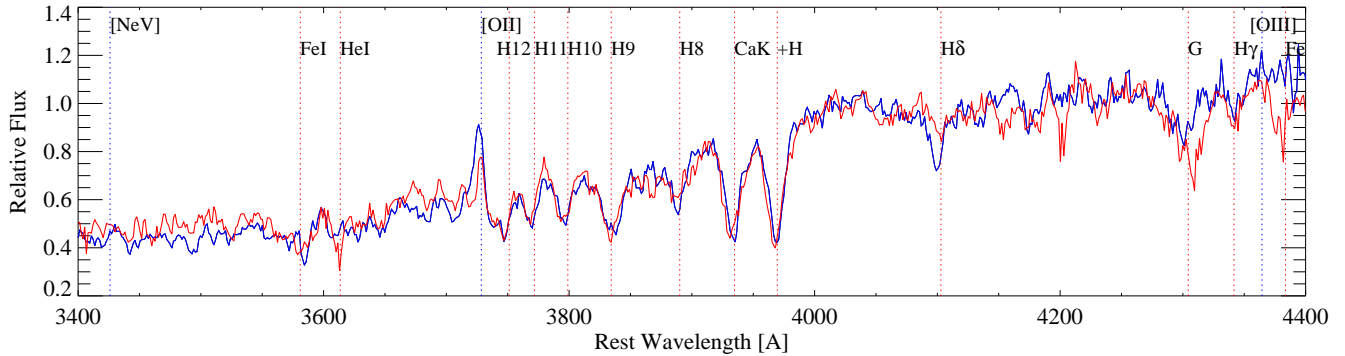}

\caption{Stacks of QGs, separated by stellar density relative to the size-mass relation: compact galaxies (red) exhibit older spectral features than more extended galaxies (blue). Absorption features are identified by red dotted lines, emission features are identified with blue dotted lines. [\ion{O}{2}] emission and H$\delta$ absorption are stronger, and the G-band is weaker, in normal QGs.} 
\label{stack}
\end{center}
\end{figure*}

To produce average composite spectra (stacks) we perform the following procedure. We first individually transform each spectrum into the rest frame using the published redshifts \citep{Vanzella2008, Popesso2009, Kurk2009}. We then flux normalize each spectrum using the median flux value measured between restframe 4000\AA $<\lambda<$4050\AA.
Finally, we stack the normalized spectra using the {\it scombine}  package in IRAF, and rebin to a common dispersion ($d\lambda = $1.4\AA  pixel$^{-1}$, rest frame), and performed a 3$\sigma$ clipping during the stack.  Our stacked composite spectra of the two QG samples are presented in Figure \ref{stack}. 
There are 6 out of the 28 compact QGs that do not have spectral coverage at red ward of rest frame 4000\AA, thus we have excluded them from the stack and its analysis presented in Section 3.2. However their spectra are suitable for the analysis presented in Sections 3.3 and 3.4, which include the full sample of 28 galaxies.

We have estimated the error on the stacks as the sample standard deviations of each spectroscopic sample. To measure the standard deviations, we repeat the stacks of each sample using jackknife resampling, each time removing one spectrum and stacking the rest of the spectra following the identical procedure described above. The final sample error is the standard deviation of the jackknifed stacks at each spectral point.

\section{Results}
\label{Results}

\subsection{Composite Spectra}

The stacks presented in Figure \ref{stack} exhibit many features typical of old stellar populations, namely strong balmer absorption typical of post-starburst galaxies, strong G-band absorption, and a prominent 4000\AA\ break. Additionally, very weak [\ion{O}{2}] emission is occasionally present (as we discuss later), indicating little (if any) star-formation in some cases. We discuss the [\ion{O}{2}] properties of the galaxies further in Section 3.3. 
In the following Section 3.2, we show the properties of the stellar populations, including average stellar population ages, in these two samples.

\subsection{Stellar Ages}

\subsubsection{Estimates from Lick Indices}

\begin{figure*} [!t]
\begin{center}
\includegraphics[scale=0.43]{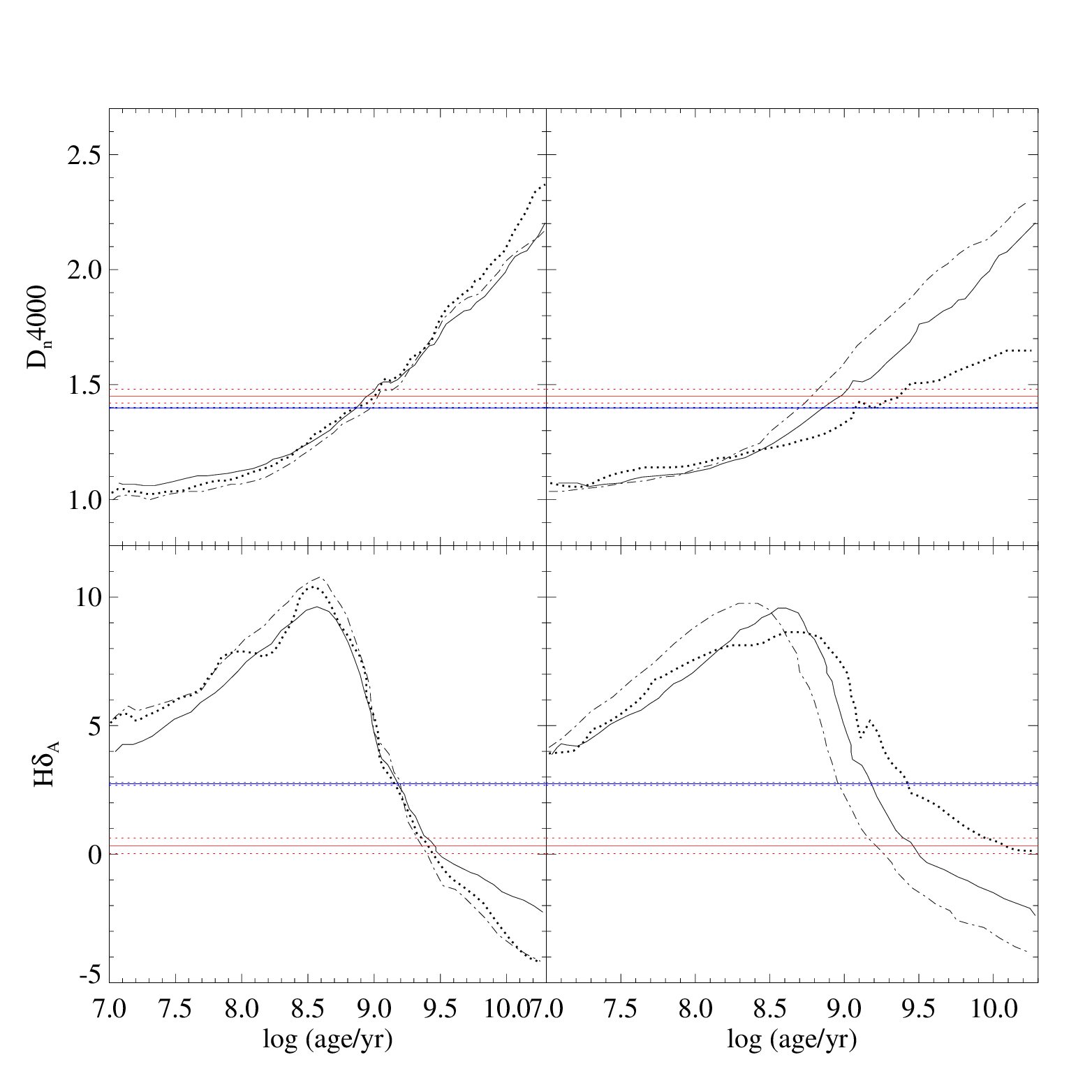}

\caption{The evolution of both age diagnostics (D$_{n}$4000 and H$\delta_{A}$) as a function of stellar age following a single instantaneous burst of star-formation using stellar population models of solar-metallicity. Left panels show evolution of solar metallicity single-age stellar populations from the STELIB library \citep[][solid line]{LeBorgne2003}, the \citet{Pickles1998} library (dotted line), and \citet[][dashed line]{Jacoby1984}. Right: The evolution of age diagnostics in the STELIB library for bursts of solar metallicity (solid line), 20\% solar (dotted) and 2.5 times solar (dashed line). The observed age diagnostics of the two populations studied here are in red (compact) and blue (normal). Figure adapted from \citet{Kauffmann2003}.} 
\label{Kauff03}
\end{center}
\end{figure*}

We have measured the average age of the stellar populations of the two QG samples from their stacks, using two age diagnostics: the 4000\AA\ break (D$_{n}$4000), and H$\delta$ absorption. For the D$_{n}$4000 diagnostic, we have used the age calibration presented in \citet{Balogh1999, Kauffmann2003}, which takes the ratio of the mean flux between 4000\AA\ and 4100\AA, to the mean flux between 3850\AA\ and 3950\AA\ in the stack. We measure the error on the D$_{n}$4000 diagnostic by generating 1000 gaussian deviates of each spectral point in the stack using the observed sample error described in Section 2.3, and repeating the D$_{n}$4000 measurement each time. The error is then the standard deviation of the sample of gaussian deviated D$_{n}$4000 measurements. We find that the compact sample have a D$_{n}$4000 $=$ 1.45 $\pm$ 0.03, a larger measure (i.e. older age) than for the normal sample, from which we have measured D$_{n}$4000 $=$ 1.398$\pm$0.002 (although the difference is marginal; $\sim$1.7$\sigma$).
We have compared these D$_{n}$4000 measurements to those made with single-age stellar population models using the same procedure \citep[Figure \ref{Kauff03};][]{Kauffmann2003}. Adapted from \citet{Kauffmann2003}, this Figure (top left panel) presents the evolution of this diagnostic as a function of stellar age at solar metallicity for a single instantaneous starburst of star-formation according to the STELIB library \citep[][solid line]{LeBorgne2003}, the \citet{Pickles1998} library (dotted line), and \citet[][dashed line]{Jacoby1984}. In the top right panel, the age evolution of the diagnostic is presented from the STELIB library for bursts of solar metallicity (solid line), 20\% solar (dotted) and 2.5 times solar (dashed line). For all models, the D$_{n}$4000 of the compact sample implies older stellar age than that of the normal sample. 

Similarly, we have made the comparison using the age-sensitive H$\delta$ absorption feature. Unlike the D$_{n}$4000 diagnostic, which grows monotonically with age, the H$\delta$ (as well as other balmer) absorption feature peaks for stellar populations of age $\sim$1 Gyr (A-type stars). As the stellar population ages, the balmer absorption starts to decrease in strength. We measure H$\delta_{A}$ using the Lick absorption line index \citep{WortheyOttaviani1997}. To estimate the error on our measurement of H$\delta_{A}$ we follow the same procedure as for D$_{n}$4000, by re-measuring the index from 1000 gaussian deviates of each spectral point, and taking the standard deviation of these H$\delta_{A}$ measurements. We find that the compact sample has H$\delta_{A}$  $=$ 0.33 $\pm$ 0.31, a smaller value (i.e. older age) than for the normal sample, from which we have measured H$\delta_{A}$  $=$ 2.74 $\pm$ 0.06. 
The bottom panels of Figure \ref{Kauff03} show how these measured values compare to those from the evolutionary models explored in \citet{Kauffmann2003}. 

Although the trends between the age diagnostics and the stellar population age are obvious for each of the models shown in Figure \ref{Kauff03}, it is also obvious that the diagnostics depend on other features of the models such as metallicity. Therefore, we do not attempt to use these measurements as average age measurements of the galaxy populations. Rather, we seek to gain insight into {\it  relative} age differences between the two samples using the few models presented in Figure \ref{Kauff03}. At face value, the average age implied by the 6 models from D$_{n}$4000 suggests an age difference between the two galaxy samples of roughly $\sim$0.3 Gyr. The H$\delta_{A}$ index suggests a larger age difference, with the compact sample are $\sim$2.5 Gyr older than the normal QG sample (we discuss in Section 3.2.2 why the difference implied by H$\delta_{A}$ is likely overestimated). However, it is clear that both age indicators imply that the compact sample, on average, have older stellar ages than the normal sample. 

 We verified that the qualitative results are robust to the compactness definition by increasing it to 1.2-$\sigma$ below the z$\sim$0 mean, which corresponds to roughly 5.8x10$^{9}$ M$_{\odot}$/kpc$^{2}$, a factor of $\sim$2 denser than the definition outlined above. This splits the compact sample defined by the 1$\sigma$ line roughly in half, resulting in 16 compact galaxies and 48 extended ones. With the smaller sample there is a significant decrease in signal to noise of the compact sample with respect to the extended one, but we are able to measure age diagnostics. We find that the D$_{n}$4000 of the more stringently selected compact galaxies increases on average, suggesting an older age, although within the errors of the previous measurement. The H$\delta_{A}$ measurement in the more stringently selected compact sample increases slightly, in the sense of younger age, however it is again within the errors. We note that the error on the compact H$\delta_{A}$ measurement is larger with the more stringent selection, likely because the H$\delta_{A}$ is typically not individually detected in the spectra, and this sample is small. Both age diagnostics of the extended sample from the more stringent selection change towards older stellar age; this makes sense in the context of our interpretation, because we have essentially added 12 (formerly) compact galaxies whose average age is older to a younger sample, and the expected effect would be to increase the age. This is in fact what we see. We interpret the changes in diagnostics from the stringently selected compact sample to be consistent with this picture: the D$_{n}$4000 shows an increase in age (although consistent within the errors of the 1-$\sigma$ selected compact sample) and the noise in the H$\delta_{A}$ measurement has increased due to decrease signal to noise from the small sample.

\subsubsection{Other Spectral Features}
In the literature, it has been extensively discussed that other spectral features adjacent to the H$\delta$ absorption line can affect the measurement of the H$\delta_{A}$ Lick index \citep[e.g.][]{Dressler2004, Prochaska2007}. In particular, stellar continuum absorption by molecular CH and CN lines to the blue and red of H$\delta$ can depress the pseudo continuum regions used to measure the index, resulting in an underestimated value for the index. The net consequence is an overestimated age based on the line. Although the nature of this continuum absorption is not well understood, it is likely originating in old stellar populations, from metal enriched cool stars \citep{Schiavon2002a, Dressler2004}. 

As can be seen in Figure \ref{stack}, we observe the compact QG stack to exhibit a prominent peak immediately blue ward of the H$\delta$ line, followed by a band of continuum absorption between 4100-4200\AA. The blue peak is an indication of the true continuum level, unaffected by the CN absorption. Such prominent features are not seen in the normal QG stack. This feature is undoubtedly affecting the H$\delta_{A}$ measurement in the compact sample, such that the implied age is overestimated (and, therefore, the relative age difference as well). 

However, this evidence of old, metal enriched stellar populations in the compact sample is in general agreement with the evidence for an older average stellar age than the normal QG sample. We discuss further constraints on this age differential between the two samples in the following section.

\begin{figure*} [!t]
\begin{center}
\includegraphics[scale=0.55, trim=0 0 10 0,clip]{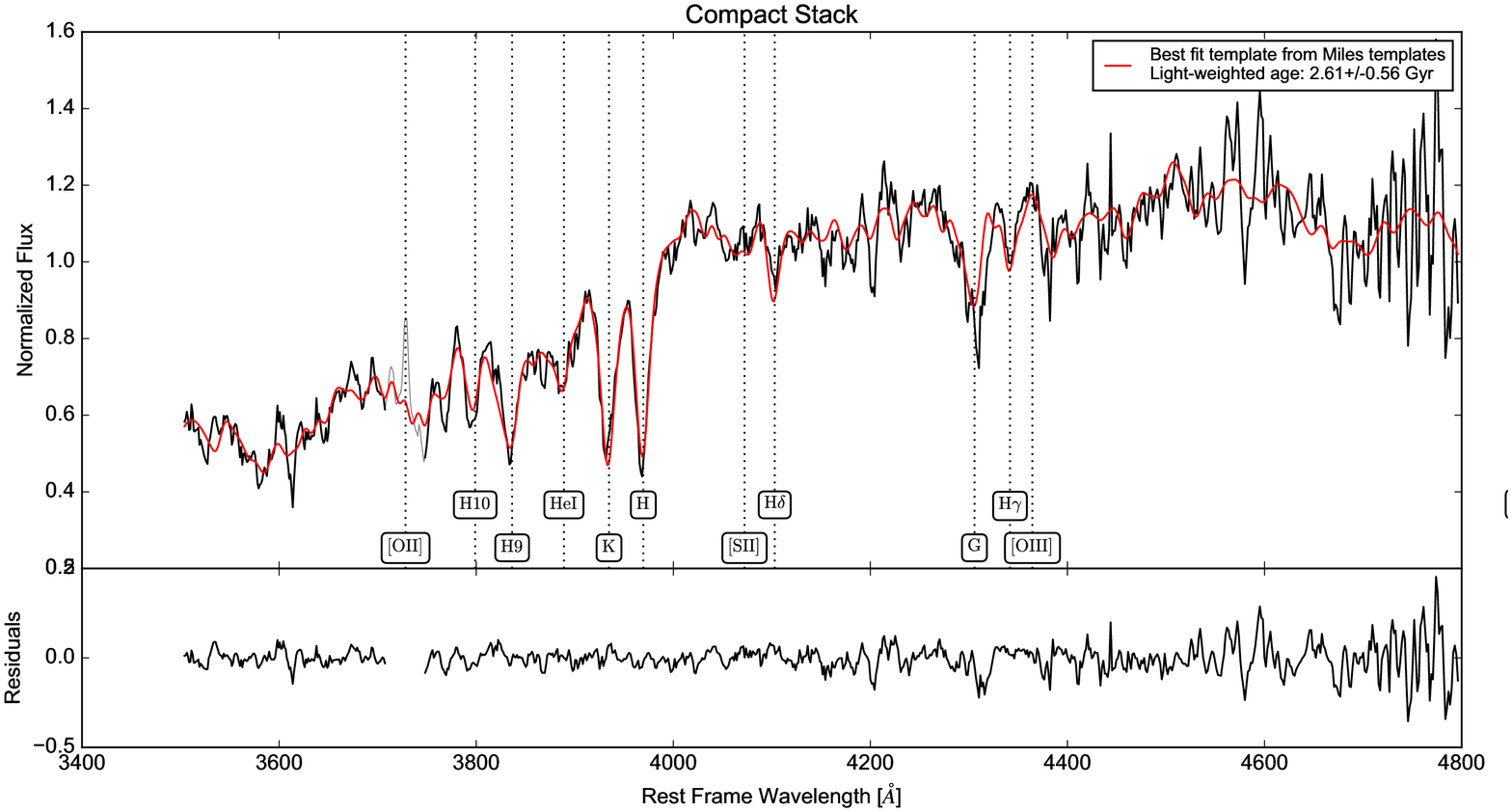}
\includegraphics[scale=0.55, trim=0 0 10 0,clip]{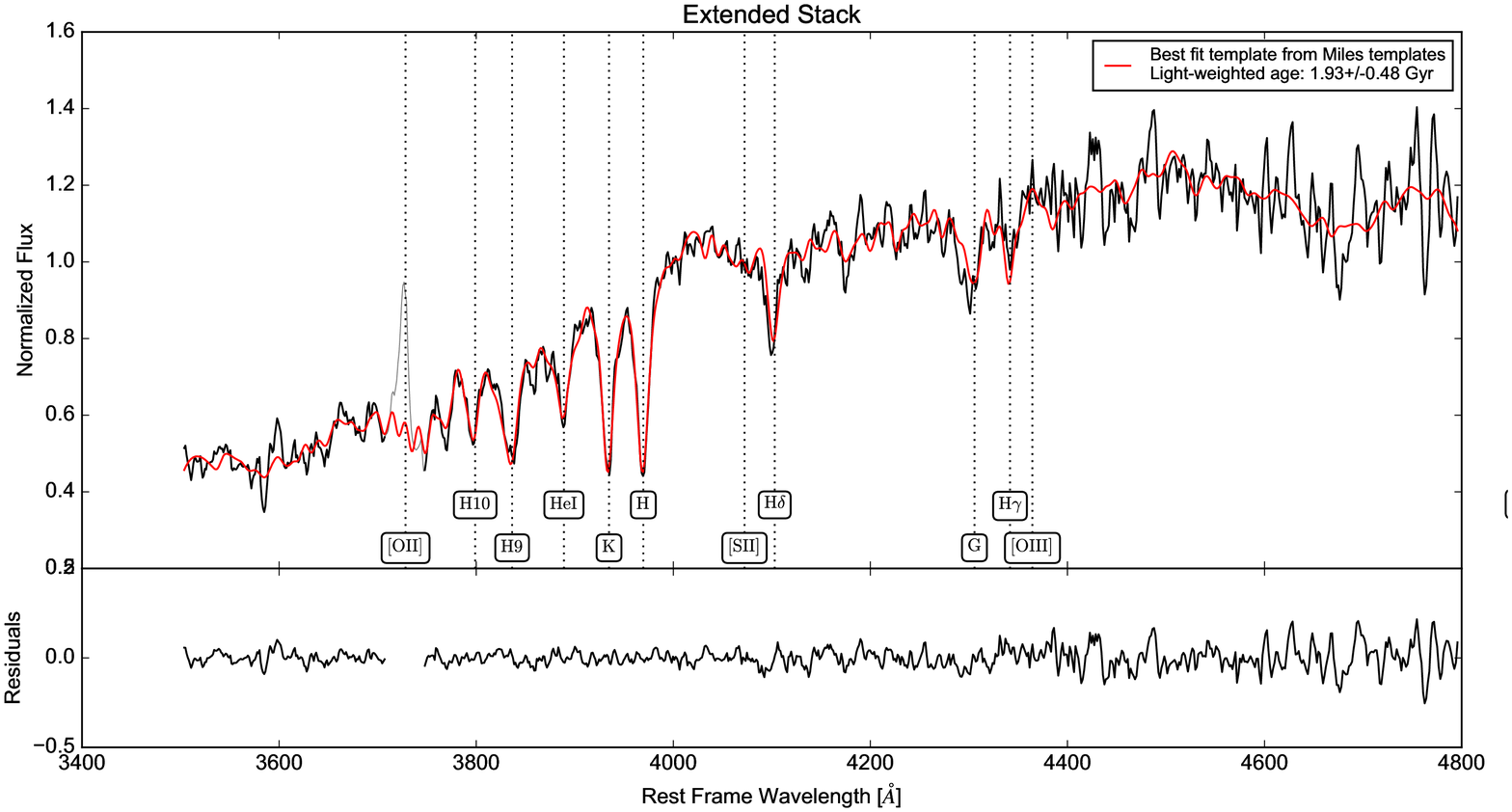}

\caption{Results from the pPXF fitting of the two stacks. Top set of panels illustrate the best fit model to the compact QG stack. The region around the [\ion{O}{2}] emission line was excluded from the fit. The bottom set illustrates the fit for the stack of normal-sized QGs. The light-weighted best fit ages between the two samples for the variety of fits indicate an older average age from the compact sample. } 
\label{ppxf}
\end{center}
\end{figure*}

\subsubsection{Stellar Population Synthesis Modeling}
An alternative estimate for the age differential between the two populations can be obtained by fitting stellar population templates to the stacks. In this section, we use the Penalized Pixel-Fitting (pPXF) Software \citep{CappellariEmsellem2004} to explore the average ages of the two galaxy samples, again with the goal of gaining insight into {\it relative} age differences. As templates for the fitting, we use the \citet{Vazdekis2010} SSP models based on the Medium resolution INT Library of Empirical Spectra (MILES; Sanchez-Blazquez et al. 2006). Due to the poorly understood nature of the CN and CH stellar continuum absorption features, we choose to use these templates because they are composed of empirical stellar spectra where these features are observed \citep[e.g.][]{Vazdekis1999}.

We allow pPXF to choose the ''optimal'' combination of templates, along with multiplicative and additive polynomials, to fit the continuum shape,  to fit each stacked spectrum, where the initial mass function (IMF) is fixed with a slope of 1.3 \citep{Salpeter1955}, and we constrain the maximum age of any template to be the age of the Universe at z=1.2. We allow pPXF to choose from a range of templates of varying ages, which are weighted and summed together to construct the best fitting model. The signal to noise of our stacks starts to deteriorate at wavelengths longer than rest-frame 4800\AA\ from a combination of decreased number of spectra with wavelength coverage in that region, and possibly poor subtraction of telluric features, and so we constrain the spectral region provided to pPXF to 3500-4800\AA, masking out 40\AA\ surrounding the [\ion{O}{2}] $\lambda$3727 emission. 

To choose the appropriate metallicity range for the fitting, we use constraints from the mass-stellar metallicity relation out to z$\sim$0.7 \citep{Gallazzi2005,Gallazzi2014} at our average stellar masses ($<M_*>\sim$10.7 for both samples). Only the solar metallicity models in \citet[][using the Padova+00 isochrones of \citet{Girardi2000}]{Vazdekis2010} fall within the confidence intervals of the \citet{Gallazzi2014} mass - metallicity relation at this average mass. Therefore, we limit our fits to these models.
 pPXF measures the luminosity weighted age based on the sum of 2 templates, which assumes galaxies may be composed of several stellar populations. Those individual populations may not necessarily follow the mass-stellar metallicity relation. However, we note that our results do not strongly depend on either the wavelength range or metallicity assumed by the fit. 

We estimate errors on the age of the stellar populations given by the fit as the variance of a series of bootstrap resampled templates added to the fit residuals. We find that the compact stack has an average age of 2.54$\pm$0.63 Gyr (reduced $\chi^{2}$=15.3), while the normal sample are 1.87$\pm$0.65 Gyr (reduced $\chi^{2}$=30.2), an age differential that is in general agreement with the findings using the Lick indices in Section 3.2.1 in that the age measured from the compact sample is older. These results are shown in Figure \ref{ppxf}.

To conclude this section, our results from both the spectral indices and the full spectral fitting {\it consistently} indicate that there is evidence that the average age of the compact QG sample may be older than that of the normal QG sample.

\begin{figure*} [!t]
\begin{center}
\includegraphics[scale=0.43]{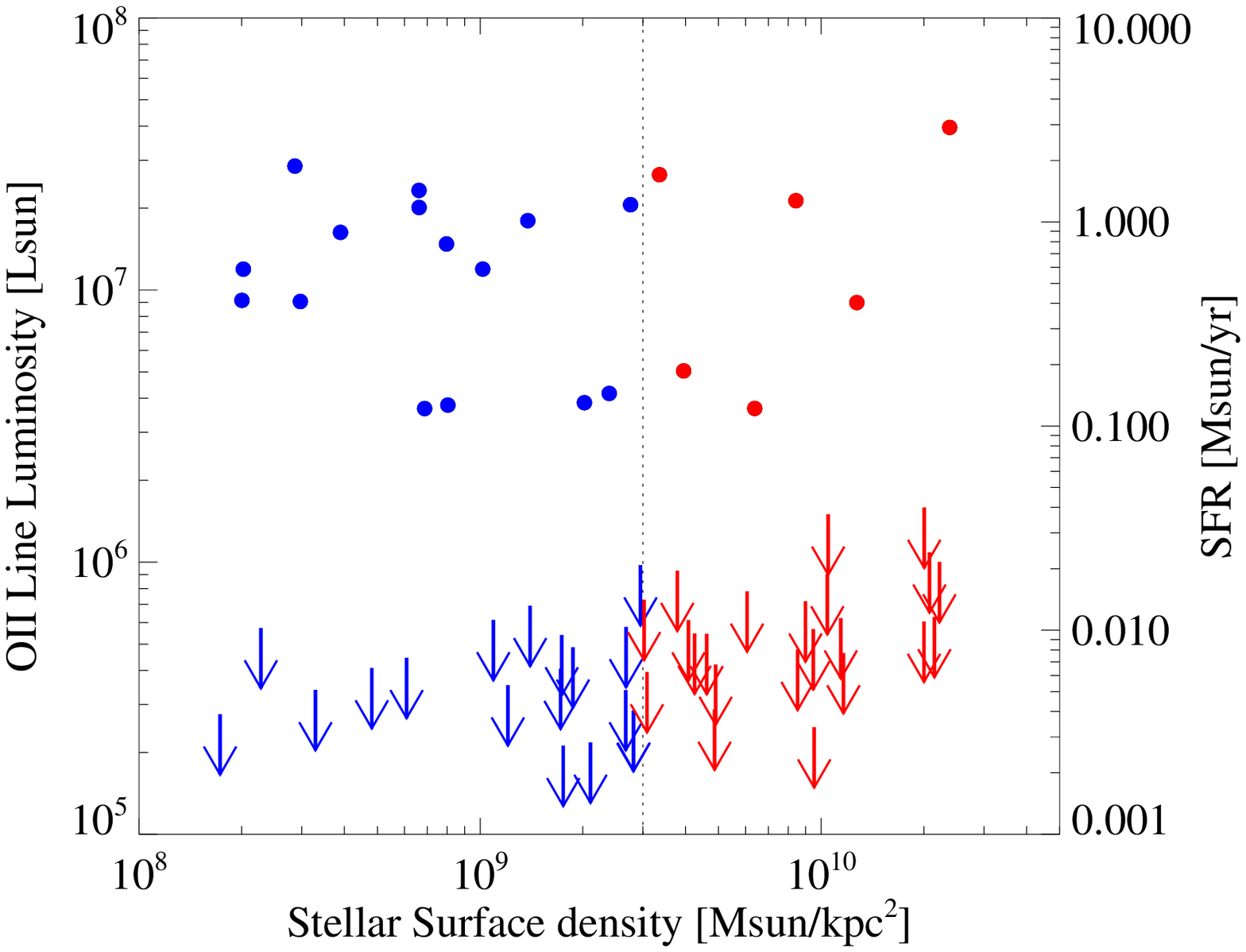}
\includegraphics[scale=0.43]{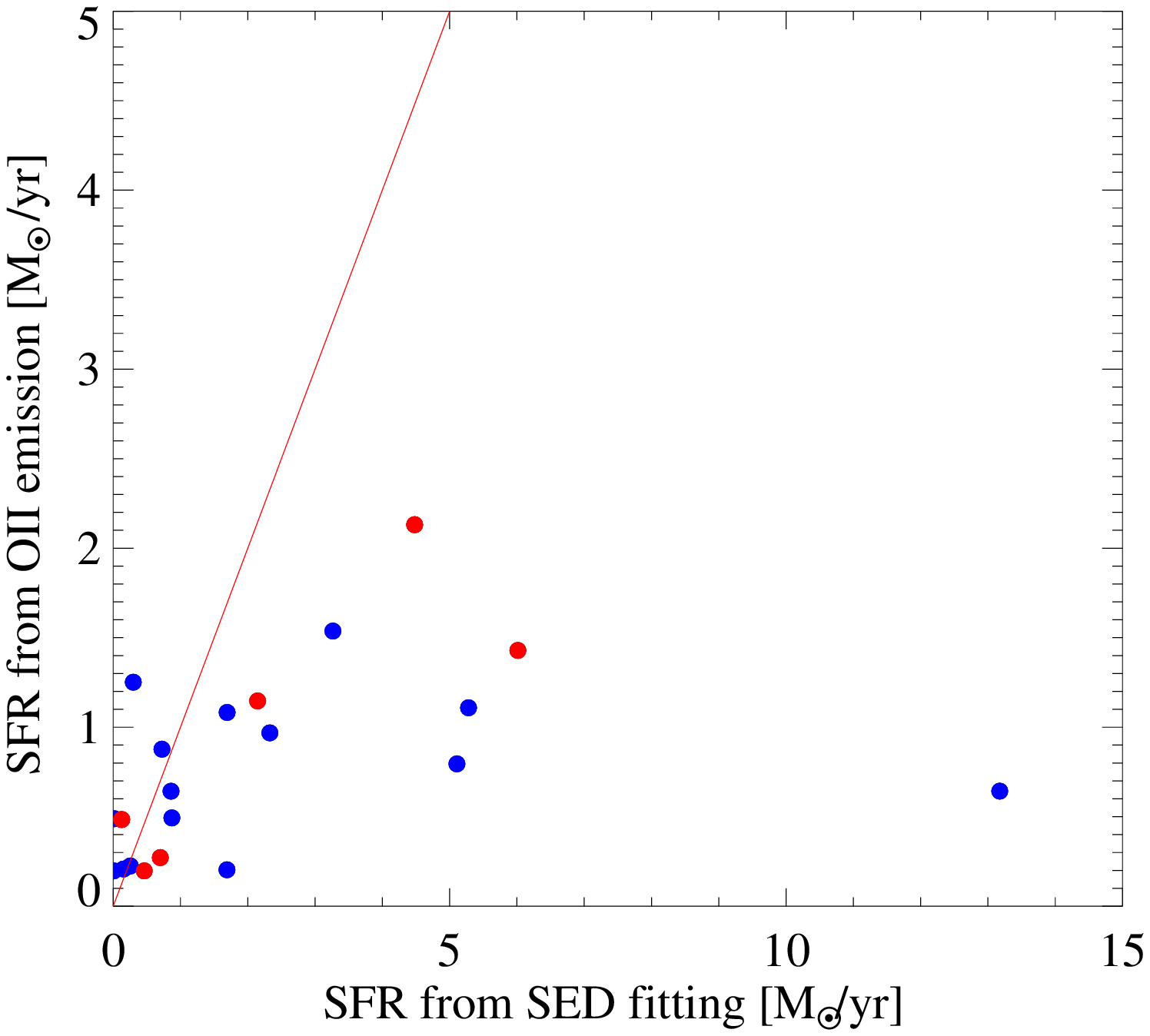}

\caption{Left: the [\ion{O}{2}] emission line fluxes (when detected) plotted vs the stellar surface density (compactness). There is no trend of flux with compactness. Upper limits for spectra without detected [\ion{O}{2}] emission are indicated by the top of the downward arrows. Right: SFR predicted from the [\ion{O}{2}] line luminosities \citep[in galaxies with \ion{O}{2} detections only;][]{Kewley2004} compared to the SFR from SED fitting (Lee et al. in prep). Few galaxies show [\ion{O}{2}]-derived SFRs in excess of that estimated from broadband photometry, therefore it is unlikely that AGN contribute significantly to the [\ion{O}{2}] luminosity. } 
\label{oiifluxdist}
\end{center}
\end{figure*}

\subsection{[\ion{O}{2}] emission}
In the stacked spectra presented in Figure \ref{stack}, we find that each QG sample exhibits weak [\ion{O}{2}] $\lambda$3727 emission. 
We have visually inspected each of the individual spectra in each sample, and found that only a fraction of the individual spectra exhibit detected [\ion{O}{2}] emission, and in general it always appears weak. The right panel of Figure \ref{oiifluxdist} shows estimated SFR from the [\ion{O}{2}] flux, assuming that the emission is produced entirely by star-formation \citep{Kennicutt1998}. The [\ion{O}{2}] flux was measured from the individual spectra using the IRAF routine {\it splot}. The [\ion{O}{2}] is always weak (corresponding to $\lesssim$2 M$_{\odot}$ yr$^{-1}$; for the majority of the sample this is less than predicted from the photometry from the best-fitting SED). It is therefore unlikely that this [\ion{O}{2}] emission is contributed to significantly by an AGN (see Section 3.4 for additional constraints). There is no obvious difference in the line luminosities or SFRs between the two samples (Figure \ref{oiifluxdist}).

However, the occurrence of [\ion{O}{2}] emission is much more frequent among the normal QG sample than the compact sample (top panel of Figure \ref{masssize}, and left panel of Figure \ref{oiifluxdist}). This is the primary reason for the stronger average emission in the normal QG stack. Out of 33 normal QGs, roughly half (15) galaxies exhibit detected [\ion{O}{2}] emission. Among the 28 compact QGs only 6 have detected [\ion{O}{2}] (21\%). 

\begin{figure*} 
\begin{center}
\includegraphics[scale=0.4]{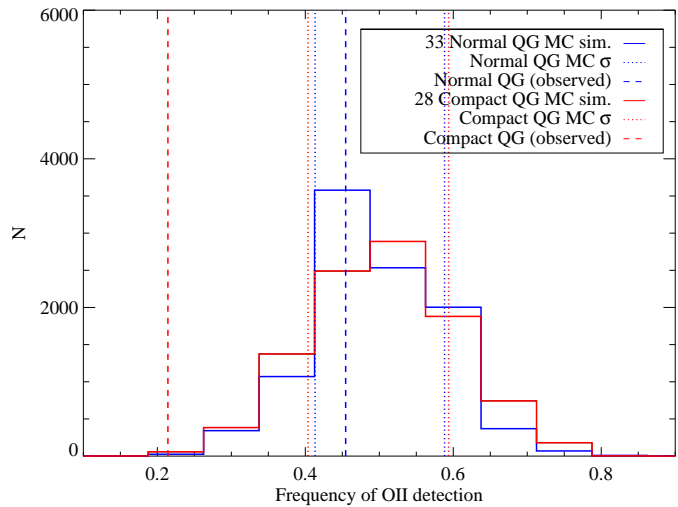}
\caption{Results of a Monte Carlo simulation to test the hypothesis that [\ion{O}{2}] detection among a sample is random. Histograms are the distribution of frequencies expected using 10000 simulated QG samples if the intrinsic detection rate is 50\% for a sample that is the same size as the compact ones (red) and the normal ones (blue). Each distribution has a mean of roughly 50\%. Standard deviations of each Monte Carlo distribution are indicated by the dotted lines. The observed frequency in the real data are the dashed vertical lines. 
Normal QGs show a detection rate consistent with a random [\ion{O}{2}] occurrence, but the compact QGs show an [\ion{O}{2}] frequency that is significantly lower.  } 
\label{OII}
\end{center}
\end{figure*}

We conduct the following analyses to assess the significance of the [\ion{O}{2}] detection rate among QGs of different stellar density. A priori we do not have reason to believe [\ion{O}{2}] emission should depend on compactness, and so we test the hypothesis that the chance of [\ion{O}{2}] emission is random; i.e. there is an equal chance (50\%) that any galaxy emits [\ion{O}{2}] as there is it doesn't. To test this hypothesis, we conduct a Monte Carlo simulation where we randomly draw samples equal to the number in each QG sample from a binomial distribution. We create 10,000 realizations, and compare the success rate (i.e. an [\ion{O}{2}] detection) to the observed [\ion{O}{2}] detection rate in each QG sample. The results of this test are presented in Figure \ref{OII}. The histograms in the top panel show the frequency of [\ion{O}{2}] detection from a sample of size 33 (normal QGs; blue) and that of a sample of size 28 (all compact QGs; red). The dotted lines show the standard deviations of the realizations. The solid lines show the observed frequency; it is clear from the figure that the normal QGs show an [\ion{O}{2}] frequency that is consistent with this hypothesis that [\ion{O}{2}] detection is random (50\% detection rate). In contrast, the compact QGs show an [\ion{O}{2}] frequency significantly lower, inconsistent at the 3-$\sigma$ level. 

\begin{figure*} [!th]
\begin{center}
\includegraphics[scale=0.35]{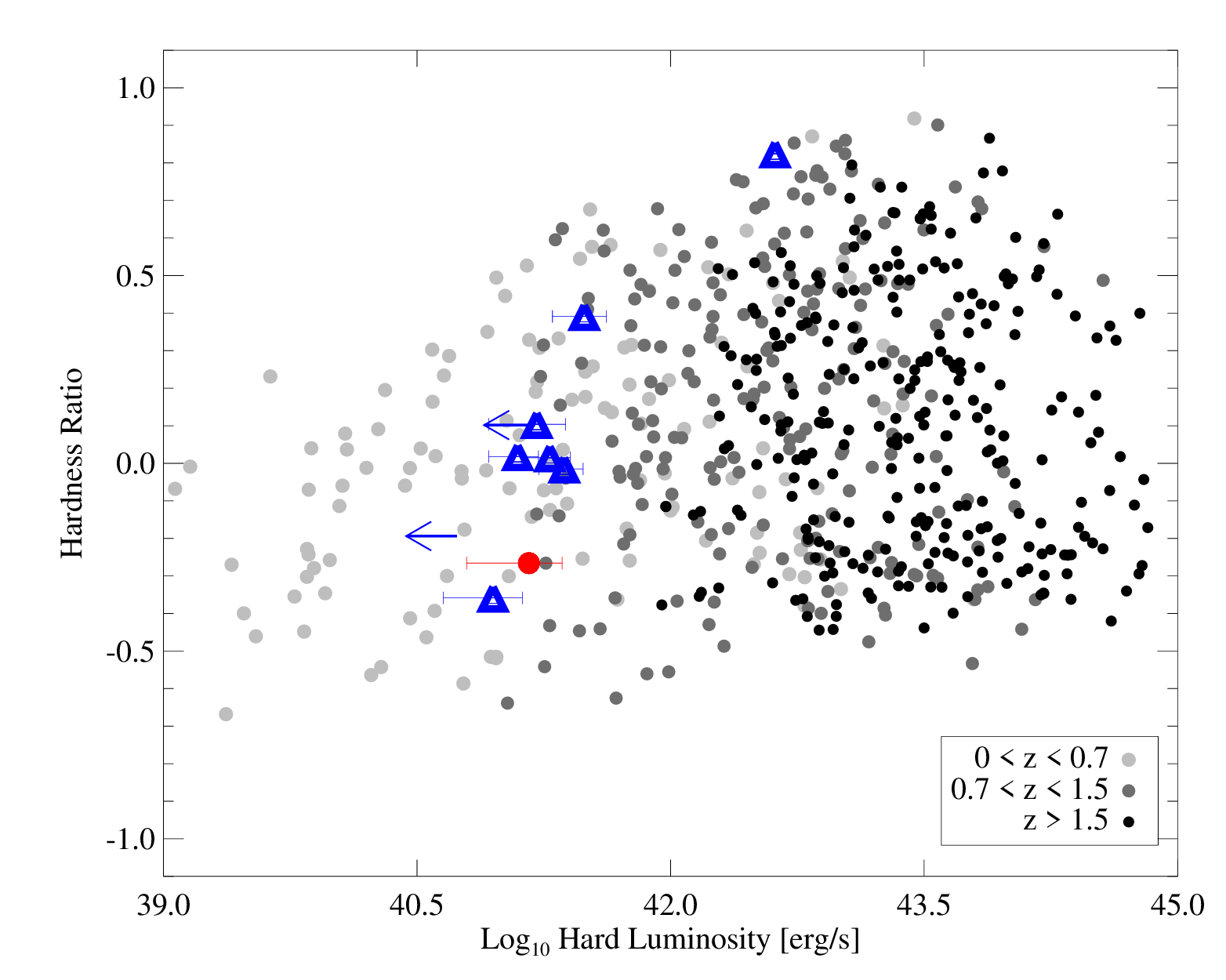}
\includegraphics[scale=0.35]{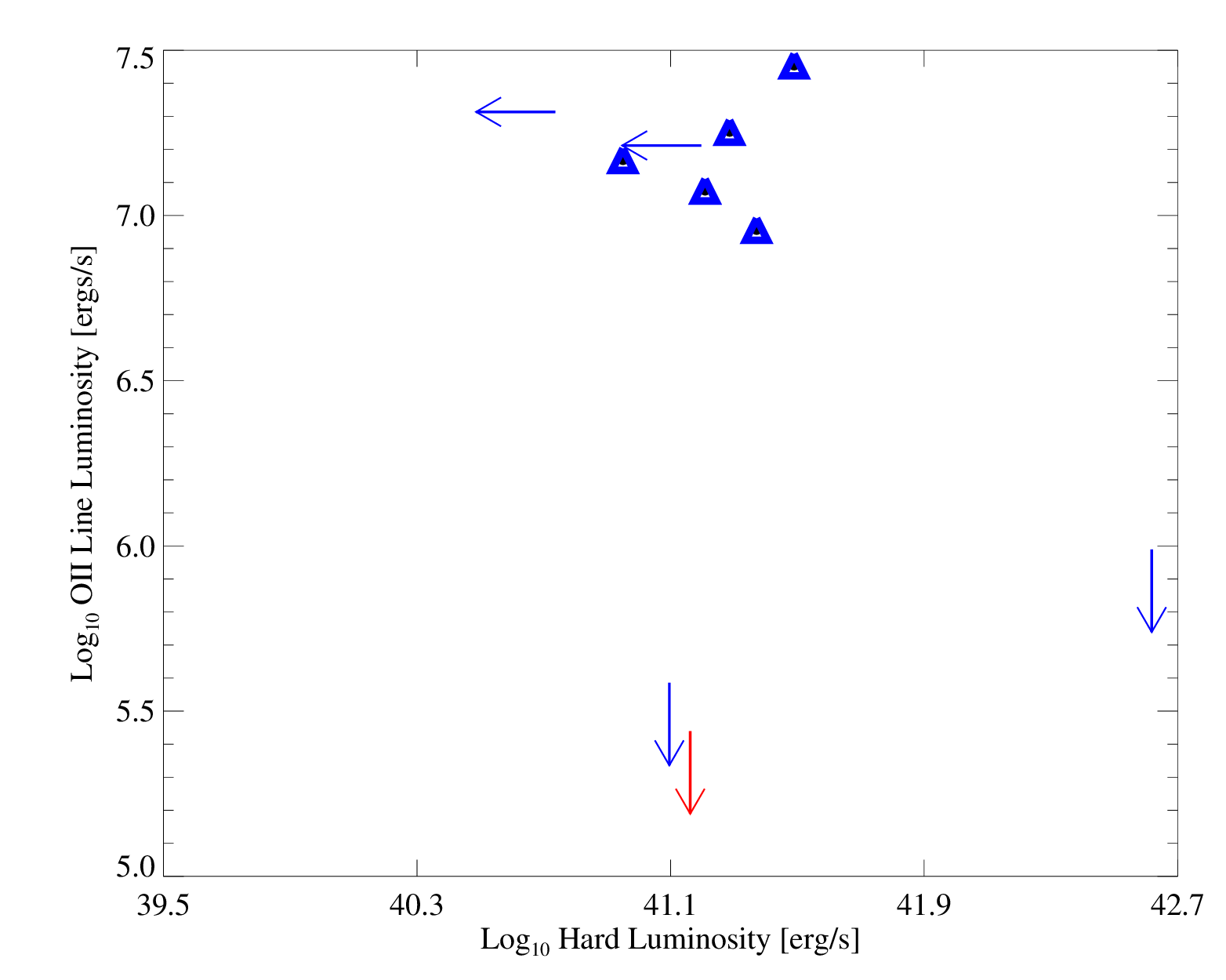}

\caption{Left: Hardness ratios vs hard-band X-ray luminosities of the detected galaxies among the two samples. Red circles are compact QG detections, and blue triangles are normal QG detections. Grey points are X-ray detections at all redshifts in GOODS-South from \citet{Cappelluti2016}.  All but one X-ray detected QG are too weak to be considered an AGN; their emission is likely due to some other source.
Right: [\ion{O}{2}] line luminosity vs hard X-ray luminosity, for X-ray detected galaxies (triangles) and upper limits to [\ion{O}{2}] luminosity when it is not detected (arrows). 
 } 
\label{xray}
\end{center}
\end{figure*}

We also investigate the significance of the differing [\ion{O}{2}] detection rate between the two samples using the Fisher exact test to measure the probability that the two QG samples are drawn from the same distribution of  [\ion{O}{2}]  emitters. We construct a 2x2 contingency table for each QG sample, and calculate the associated p-value from the Fisher exact test. We use the function {\it fisher.test} in the R statistical software environment. We calculate a 
 p-value of 0.06, indicating that we can reject the null hypothesis that the two samples come from the same parent sample of  [\ion{O}{2}] emitters. The occurrence rate of  [\ion{O}{2}]  is significantly different between the two samples at the $\sim$2-$\sigma$ level, according to this test. We conclude that the frequency of [\ion{O}{2}] detection among the compact QGs differs significantly from that of the more extended sample. Whatever produces the [\ion{O}{2}] emission in QGs, whether it is warm gas or an energy source such as residual or rejuvenated SF, it appears to be significantly less frequent in compact galaxies than in the normal sample.

\subsection{X-ray properties}

The X-ray properties of the two QG samples provides further insight into residual energy sources in these galaxies, which may or may not be related to the quenching process that shut down the star-formation, or, the origin of the [\ion{O}{2}] emission when present.
The majority of our QGs are undetected (27 out of 28 compact QGs; 
 24 out of 33 normal QGs) in the {\it Chandra}  7 Ms data in GOODS-South \citep{Luo2016}. 
For the X-ray counterpart identification, we have  made use of two catalogs; first, we have used the 4 Ms catalog by \citet{Cappelluti2016}, which is able to include significantly fainter X-ray sources than blind detections \citep[e.g.][]{Xue2011}, due to a novel technique based on prior information of the positions of optical-NIR sources.  Second, we have used the 7 Ms source catalog presented in \citep{Luo2016}. There is excellent correspondence between the two catalogs. All galaxies detected in the 4 Ms catalog are also detected in the 7 Ms data, with the 7 Ms data providing one extra detection not present in the 4 Ms data. 
 The 7 Ms X-ray detected galaxies are identified by boxes in Figure \ref{masssize}, where we have also identified those with significantly detected [\ion{O}{2}] emission. We note that among the compact sample, {\it none} of the [\ion{O}{2}] detections come from an X-ray detected galaxy. The one X-ray detection exhibits no [\ion{O}{2}] emission. In the normal sample, a large fraction (7 out of 15) of the [\ion{O}{2}] detections come from an X-ray detected galaxy. The majority of X-ray detections (7 out of 9 X-ray detections) are [\ion{O}{2}] emitters. We present the 7 Ms X-ray fluxes, hardness ratios, and [\ion{O}{2}] luminosities for all X-ray detected sources in Table \ref{xdet}. In general, it does not appear that [\ion{O}{2}] detected galaxies exhibit any obvious difference in their X-ray properties from those without [\ion{O}{2}] emission. Similarly, there does not appear to be an obvious distinction in X-ray properties between the one detected compact QG  and the normal QG detections (see also Figure \ref{xray}). 
 To estimate the upper limits to the [\ion{O}{2}] line flux, we first measure the RMS in the continuum in the vicinity of the line, and multiply by the square root of the number of pixels per resolution element (4 and 5 pixels for FORS2 and VIMOS, respectively).

We measure hardness ratios from hard and soft band fluxes for the detections as (Hard - Soft)/(Hard + Soft), where the fluxes are in counts/sec. 
In general, AGN, depending on type, can span a range of hardness ratios \citep[e.g.][]{Szokoly2004}, as do our detected QGs, 
although the luminosities of our QGs are relatively weak for AGN \citep[][]{Mainieri2002, Szokoly2004, Hasinger2008}. The left panel of Figure \ref{xray} shows that the hardness ratio of our sample increases with hard X-ray luminosity, indicating that in general an increase in hardness ratio in these samples is driven by the increase in hard-band luminosity. For comparison we have included all X-ray detections in the catalog of \citet{Luo2016},
 where detections have been gray-scaled according to their photometric redshift (or spectroscopic redshift, if available). By number, the X-ray detections from \citet{Luo2016} 
  are dominated by AGN, but at the low-luminosity end, star-forming galaxies are abundant.
 
  The significance of the X-ray detection rate among the two samples can again be assessed using the Fisher exact test as in Section 3.3. Using a 2x2 contingency table for the X-ray detection rate among the two QG samples, 
 we find a p-value of 0.015, corresponding to a significant difference between the two samples at the $\sim$2.4$\sigma$ level. 
 We also compute the significance jointly with the [\ion{O}{2}] using a 4x2 contingency table, where the rows represent the number of galaxies in each sample which have both X-ray and [\ion{O}{2}] detections, [\ion{O}{2}] detections only, X-ray detection only, and finally no detections. We find a p-value of 0.02 from the Fisher exact test, indicating the samples differ at the $\sim$2.3-$\sigma$ level, consistent with the findings from assessing the [\ion{O}{2}] and X-ray detection rates individually.

For the non-detected sources, we stack the 7 Ms X-ray images at the position of the optical-NIR sources to gain insight into the average X-ray properties of the two samples using the {\it Chandra} stacking analysis tool CSTACK\footnote[1]{http://lambic.astrosen.unam.mx/cstack\_v4.3} v4.3.  Detected galaxies 
 were excluded from the stack. 
We have removed one compact QG from the stack due to its proximity to a very off-axis, bright X-ray source, where we suspect that leaking flux after source removal using the PSF has affected the counts in our stack. The results of the stacking analysis is presented in Table \ref{xstack}. We find significant stacked flux in the soft band for both samples, but both samples are essentially undetected in the hard band (the normal sample has a marginal detection with signal to noise $\sim$1.4). The stacked soft fluxes for both samples do not differ significantly from each other. The luminosities, assuming the average redshifts in each sample, are very low level, inconsistent with the presence of powerful AGN.  
The stacked soft-band flux (and lack of hard-band detection), which represent the average of the majority of our samples, seem to indicate that the X-ray emission is soft. Sources of soft X-ray emission in non-starforming galaxies include Low Ionization Nebular Emission Regions \citep[LINER;][]{Heckman1980}, as well as bremsstrahlung emission from hot ($\sim$1 keV) ISM or halo gas. Such hot gas emission at the luminosities we observe here have been observed in local QGs \citep[e.g.][]{Fabbiano1992, Boroson2011, KimFabbiano2013}. We discuss this possibility in more depth in Section 4.3.

\begin{table*}[!!t]
\begin{center}
\caption{ Properties of X-ray detected galaxies in 7 Ms Chandra data}
\begin{tabular}{lllllllll}

\tableline\tableline
Galaxy Sample & ID\footnote{X-ray properties from catalog published in Luo et al. 2016} & Soft flux \footnote{Soft and hard fluxes in 1e-17 [ergs/s/cm$^{2}$]. No uncertainties indicate an upper limit on the flux.} & Soft Luminosity \footnote{X-ray Luminosities in [erg/s]. No flux uncertainties indicate the luminosity is an upper limit.} & Hard flux & Hard Luminosity & Hardness Ratio & [\ion{O}{2}] Luminosity\footnote{[L$_\odot$]} &  \\
\tableline
\vspace{1mm}

Compact &     559&      2.16$^{+     0.68}_{-     0.55}$ &  9.07E+40 &      3.46$^{+     1.98}_{-     1.58}$ &  1.45E+41 &     -0.27 &  -99 \\ 
\tableline
\vspace{1mm}
Normal &     515&     1.74 & 3.28E+40 &      8.60$^{+     4.12}_{-     3.54}$ &  1.62E+41 &      0.10 &  1.19E+07 \\ 
\vspace{1mm}
&     555&      3.21$^{+     1.06}_{-     0.88}$ &  4.37E+40 &    11.56 & 1.57E+41 &      0.10 &  1.63E+07 \\ 
\vspace{1mm}
&     616&      2.69$^{+     0.78}_{-     0.63}$ &  3.73E+40 &      9.00$^{+     2.93}_{-     2.40}$ &  1.25E+41 &      0.02 &          -99 \\ 
\vspace{1mm}
&     745&      3.64$^{+     0.87}_{-     0.73}$ &  7.13E+40 &      4.55$^{+     2.24}_{-     1.84}$ &  8.91E+40 &     -0.36 &  1.48E+07 \\ 
\vspace{1mm}
&     861&      3.95$^{+     1.00}_{-     0.87}$ &  5.66E+40 &     13.49$^{+     4.33}_{-     3.88}$ &  1.93E+41 &      0.02 &  1.80E+07 \\ 
\vspace{1mm}
&     881&     1.69 & 2.91E+40 &     17.86$^{+     6.29}_{-     5.73}$ &  3.09E+41 &      0.39 &  2.85E+07 \\ 
\vspace{1mm}
&     448&      2.88$^{+     1.18}_{-     1.06}$ &  7.18E+40 &    166.27$^{+    14.24}_{-    13.58}$ &  4.15E+42 &      0.82 &          -99 \\ 
\vspace{1mm}
&     574&      2.36$^{+     0.76}_{-     0.61}$ &  3.24E+40 &     3.97 & 5.45E+40 &     -0.19 &  2.06E+07 \\ 
\vspace{1mm}
&     594&      5.14$^{+     1.15}_{-     1.02}$ &  7.42E+40 &     16.30$^{+     4.76}_{-     4.30}$ &  2.35E+41 &     -0.01 &  9.08E+06 \\

 \tableline
\end{tabular}
\label{xdet}
\end{center}
\end{table*}

\section{Discussion And Conclusion}

There are three main results in this study: 1) in the redshift
range that we have considered, $1\le z\le 1.4$, massive compact QGs 
have stellar populations that are, on average, older than normally
sized ones; 2) the frequency of [\ion{O}{2}] detection and X-ray detection are significantly
lower among the compact galaxies; and 
3) the X-ray properties generally disfavor the presence of strong AGN in both samples of recently QGs (low luminosities $\approx 10^{40}-10^{41}$ erg/sec from both the few X-ray detections, and average stacked emission). 
While X--ray detected normal QGs often also have [\ion{O}{2}] 
emission, not a single compact galaxy with [\ion{O}{2}] emission
is individually detected in Chandra images. 
This strongly argues against AGN as the power source of the  [\ion{O}{2}] 
emission in compact galaxies, and favors instead either warm gas, stellar remnants and/or
residual star formation, minor merging with a gas rich companion, or LINER emission, possibly powered by stellar sources
 \citep[e.g.][]{YanBlanton2012, Singh2013}. These mechanisms may also be active in the normal
galaxies. Although it is unclear what are the 
  sources of [\ion{O}{2}] emission, it is 
clear that they are less active among the compact sample.
Taken all together, these lines of evidence paint a picture in which
compact galaxies formed and evolved earlier than normal ones and,
consequently, quenched star-formation earlier.

\begin{table*}[!!th]
\begin{center}
\caption{ Stacked 7Ms X-ray Fluxes \label{table1}}
\begin{tabular}{llllll}

 &  & & \\

\tableline
 Flux [ergs/s/cm$^{2}$]\footnote{Energy conversion factors evaluated using the Chandra PIMMS tool  (http://cxc.harvard.edu/toolkit/pimms.jsp)}   & Galaxy Sample & Soft band  & Hard band &  \\ 
\tableline
 &  & & & \\ 

& Compact  & 3.46$\pm$1.37x10$^{-18}$&-0.30$\pm$1.73x10$^{-17}$  \\ 

& Normal  & 5.99$\pm$1.53x10$^{-18}$&2.48$\pm$1.68x10$^{-17}$  \\ 
\tableline
\tableline
Luminosity [ergs/s]\footnote{assuming the average redshift of each sample}& Galaxy Sample & Soft band  & Hard band &  \\
\tableline
&  & & \\ 

& Compact & 2.96$\pm$1.17x10$^{40}$&-0.26$\pm$1.48x10$^{41}$  \\ 

& Normal  & 4.23$\pm$1.08x10$^{40}$&1.75$\pm$1.19x10$^{41}$  \\ 
\tableline

\tableline
\end{tabular}
\label{xstack}
\end{center}
\end{table*}

\subsection{Age constraints: evidence for progenitor bias}

The evidence of the age difference between normal and compact QGs
we have presented here comes from two independent age diagnostics, the
$D_{n(4000)}$ and the H$\delta_{A}$, as well as stellar population synthesis modeling. The $D_{n(4000)}$ is larger in compact
galaxies than normal ones, 
corresponding to an age difference of $\sim$0.3 Gyr, using the calibrations by
\citet{Kauffmann2003}. The difference in H$\delta_{A}$ implies a larger age
differential ($\sim$2.5 Gyr) but as discussed in Section 3, may be somewhat
overestimated due to continuum absorption in old, metal-rich stars.  Although
these age conversions are model dependent, the evidence for an age {\it difference} between samples
is independent of the adopted stellar libraries and assumed metallicity for
the range of values found here. Stellar population synthesis modeling with
pPXF provide consistent results with the Lick Indices.
All age diagnostics considered here imply age differentials in the same direction, i.e. more
compact passive galaxies are older.

The age differential between normal and compact QGs that we
discuss here is at redshift $z\sim 1.2$. However, the result is in qualitative
and quantitative agreement with the observations by \citet{Belli2015} at
$z\sim 2$ that at a given redshift, the largest galaxies (in radius, although related to stellar density) are among the youngest, 
suggesting that the property is a general feature of passive,
massive galaxies at high redshift. Additionally, \citet{Saracco2009} have arrived at the same result via the opposite analysis from us; by separating the oldest QGs during the epoch 1$<$z$<$2 from the youngest ($\delta$age$\sim$1.5-2Gyr) they find the youngest to reside on the local z$\sim$0 early-type galaxy mass-size relation (like our normal QG sample), whereas the old QGs are denser, with radius a factor of 2.5-3 smaller than local early-type galaxies. Although, they note that their QG samples differ in mass with the younger sample being less massive \citep[see also][]{Thomas2010, Fagioli2016}. The fact that we observe the same trend with essentially an identically mass-matched sample indicates that mass is not the primary factor related to the age differential in the population, and rather the stellar density or size may be the primary factor \citep[][find a similar result]{Saracco2011}. In a complementary study, \citet{Fagioli2016} find that at lower redshifts than our sample (0.2$<$z$<$0.8) this trend of age and compactness in QGs persists in the stellar mass range explored here (however, \citet{Trujillo2011} do not find evidence for such a trend to z$\sim$0).

At high redshift, compact galaxies dominate the population of QGs
at the high mass end \citep[][]{Cimatti2008, vanDokkum2010, Cassata2011, Cassata2013, vanderWel2014}. This has prompted investigations of scenarios where quenching 
 is more efficient in galaxies with high stellar density
because of increased stellar feedback \citep[e.g.][]{Hopkins2010}. However, a causal
relationship between high stellar density (compactness) and likelihood of
quenching does not directly predict an age--density correlation; rather, at
any given epoch, the densest galaxies should be the most likely to quench, 
but not necessarily the oldest \citep[see e.g.][]{Whitaker2012, Yano2016}.  

The presence of a relationship between age and stellar density where by denser
galaxies are older is precisely the prediction of the progenitor bias scenario
\citep{LopezSanjuan2012, Carollo2013, Poggianti2013, Belli2015, Keating2015, Wellons2015, Wellons2016, LillyCarollo2016}.
  That is, due to the observed size-evolution of star-forming galaxies \citep[e.g.][]{vanderWel2014}, the density of a galaxy reflects the density of the Universe when the
galaxy formed (assuming very little, or average, structural disruption) and
therefore older galaxies should be denser. Thus, at face value, our results
support that galaxies which form earlier, and completed their evolution
earlier, were simply denser than larger galaxies that form and evolve later,
without the density necessarily having anything directly to do with the
cessation of their star formation activity. Star-formation may be then affected by some other quenching agent \citep[e.g. halo or mass quenching;][]{BirnboimDekel2003,DekelBirnboim2006, Peng2010}. 
This idea is extensively discussed in \citep{LillyCarollo2016} who show by means of a simple toy model that this scenario will naturally explain the correlations between galaxy structure and star-formation properties, without the need of a stellar density-related quenching mechanism \citep[see also][]{Abramson2016}.

\subsection{Energy sources: quenching agents in QGs}

An independent, but complementary, piece of information comes from the X--ray
and [\ion{O}{2}] properties of our two samples. In themselves the [\ion{O}{2}]
emission line and the X--ray data do not provide any firm indication as to the
causes of quenching. The average X--ray luminosity for the majority of our
sample does not show evidence of any powerful AGN, but one could have been
present prior to reaching the current, very low level of star formation
activity. 

There is evidence
that the sources of ionizing radiation or warm gas that are still present in the two
samples are different at the time of observation; namely the compact galaxies
show a significantly lower detection rate of [\ion{O}{2}] and of
X-ray emission than the normal QGs. This difference is fully
consistent with a scenario where the quenching occurred earlier in the compact
sample and have therefore had a longer time to fade. 
 Larger QGs from the normal sample are more likely to exhibit
emission from energizing sources simply because on average, quenching in
larger galaxies was initiated more recently. Alternatively, the presence of [\ion{O}{2}] emission may be evidence of rejuvenated SF due to gas from minor merging \citep[e.g.][]{Treu2002}.
In general, our data do not provide any conclusive constraints on the
quenching mechanisms that truncated the star-formation in these galaxies, nor
if the quenching mechanisms differ with stellar density.

It remains an important goal of galaxy evolution to understand the quenching
mechanisms in massive galaxies. Over the last several years, efforts have been
made to identify star-forming progenitors of soon-to-be QGs at z$>$2. These
efforts have relied on the fortuitous observation that the first galaxies to
quench are compact, making it relatively easy to identify their immediate
star-forming progenitors among compact star-forming galaxies \citep[][]{Williams2014, Barro2013, Patel2013, Stefanon2013, Nelson2014,  vanDokkum2015}. It is interesting to note that various studies have come to
very different conclusions about the nature of feedback in compact
star-forming galaxies. \citet{Barro2013} have claimed that the AGN fraction in
compact SFGs may be up to 50\%, suggesting that the presence of AGN may
truncate the star-formation on short timescales. However, \citet{Spilker2016}
 have discovered that a subset of this sample, despite being on
the star-forming main sequence, have much lower molecular gas masses than
normal main-sequence galaxies, suggesting that star-formation will be quenched
on short timescales due to simple gas exhaustion if the influx of gas has been
suppressed. Additionally, \citet{Williams2015} found no evidence for AGN among
compact star-forming galaxies, but instead detected both faster outflow
velocities in the ISM, and extreme, redshifted Lyman-$\alpha$ emission among
compact star-forming progenitors.  \citep[Similar Ly$\alpha$ signatures were
identified among quenching galaxies by][]{Taniguchi2015}. They
interpreted these observational signatures as related to the compact galaxies
having enhanced feedback in the ISM, due to higher surface density of
star-formation than their more extended counterparts \citep[see
  also][]{Alexandroff2015,DiamondStanic2012,Sell2014,HeckmanBorthakur2016},
plausibly leading to the truncation of future star-formation.

From our data it is clear, however, that in the time since quenching was
initiated in these QG samples, major energizing sources have already
dissipated. Future detailed studies of galaxies closer to their quenching
epoch may identify the physical processes that shut down star-formation at
high-redshift.

\begin{figure*} [!th]
\begin{center}
\includegraphics[scale=0.35]{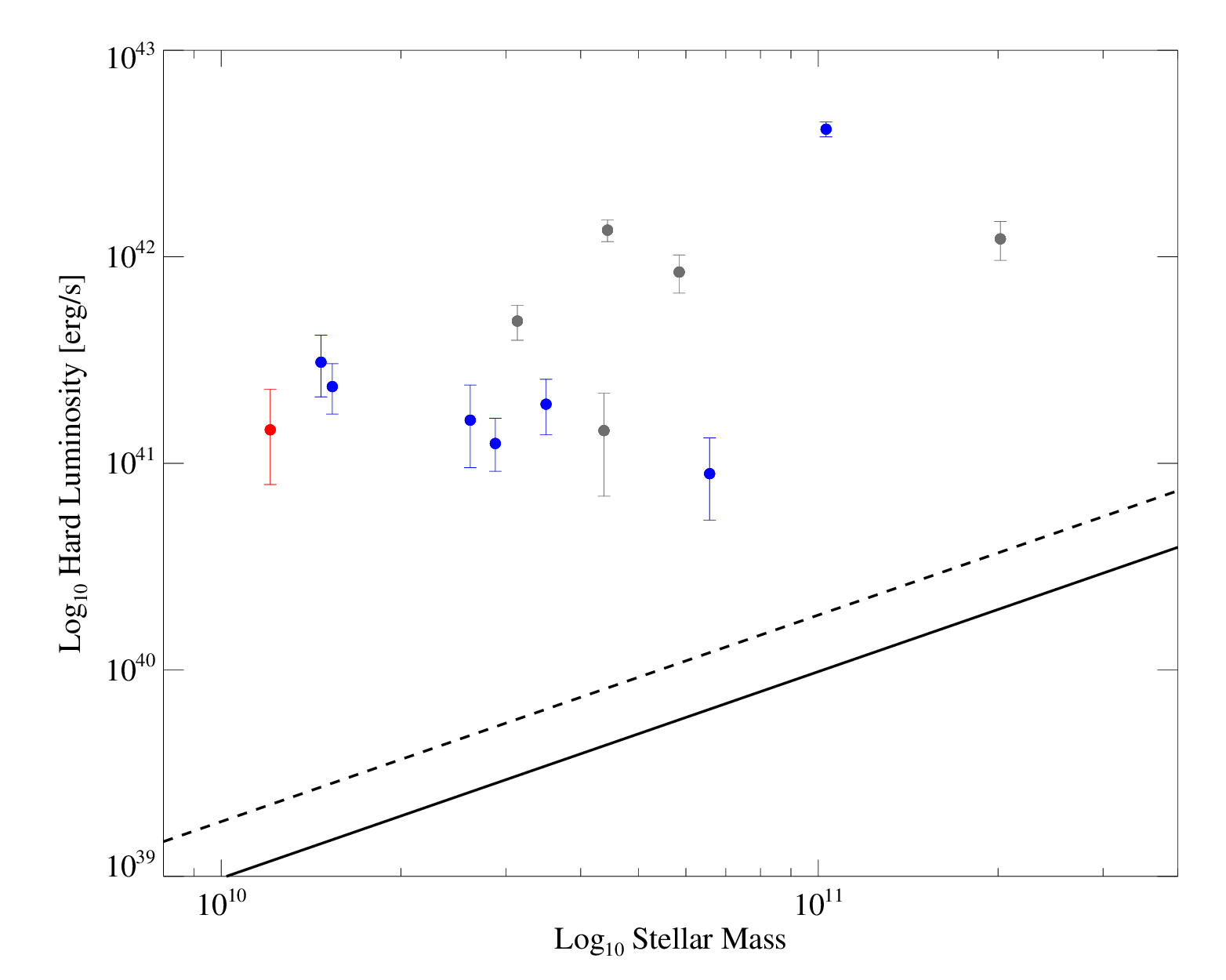}
\includegraphics[scale=0.35]{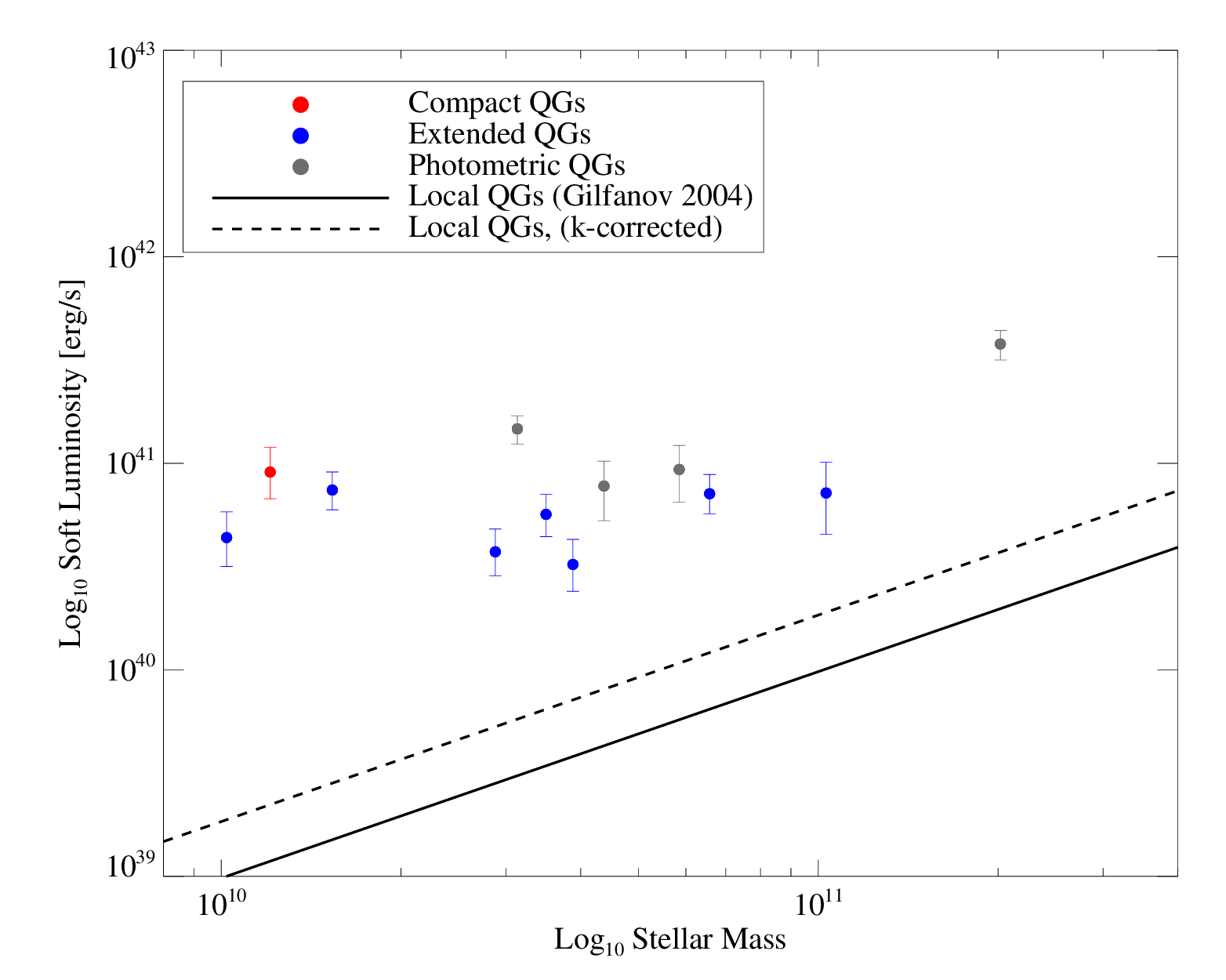}

\caption{Left panel: 7 Ms hard-band detected QGs, including photometric QG without optical spectroscopy (gray points) compared to the local QG relation for total X--ray luminosity (0.3-10 keV) from LMXB \citep[solid black line;][]{Gilfanov2004}. We have additionally k-corrected the local QG relationship to z$\sim$1.2 (dashed line) assuming a spectral slope as described in the text. Right panel: The same relation for the 7 Ms soft-band detected QGs. 
  The X--ray luminosity from the z$\sim$1.2 QG sample is larger than that emitted purely by LMXB in local QGs at a given stellar mass, plausibly suggesting additional sources of X-ray emission. }
\label{xmass}
\end{center}
\end{figure*}

\subsection{X-ray emission from passive galaxies: AGN, binaries, or hot gas?}

An important feature of both samples is that some of the QGs have X-ray
emission detected in the deep Chandra 7 Ms data of GOODS-South. 
AGN have long been suspected to be the key agent behind the quenching of star
formation \citep[e.g.][]{Granato2004}, as well as prevention of future star formation in galaxies \citep[e.g.][]{CiottiOstriker1997, Fabian2012}. Thus an
obvious question is whether or not the X-ray properties of our QGs are
consistent with this idea. 
Only a minority of the galaxies are individually detected (see
Table 1); stacking the images at the position of the the non-detected
galaxies, however, yields measurable flux in the soft band, but not in the
hard one.  
While the range of hardness ratio spanned by our
galaxies is the same as narrow--line AGN and normal galaxies \citep[e.g.][]{Szokoly2004, Hasinger2008},
the distribution of hardness ratio and X-ray luminosity shown by our galaxies qualitatively looks
very different from the distribution of the parent population of X-ray
detections \citep[see gray points in Figure 
\ref{xray};][]{Luo2016}. In 
number, these X-ray sources are dominated by AGN with luminosities $>$
10$^{42}$ erg/s \citep[see also][for the distribution among more powerful
  obscured AGN]{Wilkes2013}. The distribution exhibits large scatter in
hardness ratio and X-ray luminosity, most likely a reflection of the diversity
of obscuration, orientation, spectral slope, power and selection effects of
the AGN population.  Our X--ray detected sample have significantly lower luminosities than the majority of AGNs at comparable redshifts, suggesting a different origin for the X--ray emission (with the exception of the brightest QG, whose high X--ray luminosity and extreme hardness ratio suggest it may host an obscured AGN).

 Figure 7 also shows a plot of the X-ray luminosity 
 versus the \ion{O}{2}\ luminosity
(right panel). We do not find a correlation between X-ray and \ion{O}{2}\ luminosities, which, if present, may have suggested an AGN origin for both (however we note the sample has limited dynamic range in \ion{O}{2}\ luminosity).  These arguments, taken together with the fact that we observe stacked X-ray emission in the soft band but not the hard one (i.e. on average our galaxies are soft X-ray sources) suggests that the X-ray emission is unlikely powered by AGN.

Alternatively, the X--ray luminosity of our QGs could be primarily powered by
emission from hot gas (e.g. coronal gas that formed as a combination of
outflows, stellar winds, or gravitational heating), and/or low--mass X--ray binaries (LMXB). 
To investigate the contribution from hot gas and LMXB,
 we plot the relationship of both the hard and soft X--ray luminosity with stellar mass for our X--ray detected samples in Figure \ref{xmass}. At z$\sim$0, it is well established that X--ray luminosity from LMXB scales linearly with stellar mass \citep[e.g.][]{Gilfanov2004, KimFabbiano2004,Colbert2004}. 
 We find that our spectroscopic sample (colored points) does not show any correlation between X--ray luminosity and stellar mass, as would be expected if the X--ray luminosity was primarily from LMXBs. 
Since our X--ray detected sample is small, we additionally plot photometrically selected QGs from the parent sample selected in \citet{Cassata2013} at 1$<$z$<$1.5 that are X--ray detected in the {\it Chandra} 7 Ms imaging (i.e. those that were excluded from this study due to lack of optical spectroscopy). We limit the photometric QG sample to this narrow redshift range to mitigate the uncertain effects due to comparing the luminosities without the k-correction, which becomes increasingly large as the redshift range within the sample increases. 
The X--ray luminosities we observe in our QG sample (both hard and soft) are well in excess of that seen from LMXB in local QGs \citep[solid line in Figure \ref{xmass};][]{Gilfanov2004}.
The local scaling relation appears an order of magnitude lower than our data, despite the luminosity being the total sum of photons with energies spanning both the hard and soft bands \citep[0.3-10 keV;][]{Gilfanov2004}. 
We k-corrected the local QG relation to the observed-frame of the QG sample (with mean redshift z$\sim$1.2) assuming a conservative (steep) but typical power law spectral shape defined by a photon index of $\Gamma \sim$1.8, typical for low-redshift LMXB \citep{Lehmer2007, Boroson2011}. The k-correction increases the expected luminosity at z$\sim$1.2, however, is still well below our observations. 

Observationally it is unknown if the X--ray luminosity from LMXB evolves with redshift; although theoretical scaling relations 
 indicate the possibility that LMXB luminosity may increase with redshift, at a given stellar mass, based on metallicity and star-formation history evolution in the Universe \citep{Fragos2013}. We are unable to place further constraints on the contribution of LMXB. We conclude by simply noting the following: 1) our QGs are an order of magnitude more luminous in the X--ray than the LMXB contribution in local QGs of the same mass, however, 2) their observed X--ray luminosities are comparable to the total emission of local massive QGs, which includes the luminosity emitted from hot gas \citep[e.g.][]{KimFabbiano2013, Goulding2016}. 
Locally, it is known that the contribution from $\sim$1 keV gas dominates the soft-band emission \citep{Matsumoto1997, Sivakoff2004, Gilfanov2004}, also consistent with the luminosity seen on average in the stacked z$\sim$1.2 QGs.
This leaves open the likely possibility that some X--ray emission comes from hot gas in these QGs, heated either by gravitational or feedback processes. However, the dominant X--ray source in these z$\sim$1.2 QGs remains unknown. 

An in--depth study of the X--ray emission of our sample and the comparison
with local counterparts is beyond the scope of this paper, which is devoted to
the galaxies' stellar age. We simply note that the possibility that we are
observing hot gas emission from individual quenched galaxies at $z\sim 1.2$, an observation which is
very important to understanding the evolution of the interstellar medium and circum-galactic medium in early-type
galaxies. In a forthcoming paper, we plan to accurately quantify selection
effects and further investigate the X-ray properties of QGs.  If confirmed, it
will provide a powerful diagnostic of the feedback that took place during the
star--formation phase \citep[e.g.][]{vandeVoort2016} and that is preventing the galaxies from forming stars
again. We also note that even if the observed X--ray emission does indeed turn
out to be from hot gas, this still does not rule out AGN as the cause of
quenching, since the AGN activity could have been quenched together with the
star-formation, for example, as a result of the cessation of gas accretion
into the galaxies. The question of whether or not AGN is responsible for
star-formation quenching in galaxies thus remains an open one.
 However, if hot gas is indeed a major component of the X--ray emission, this suggests that the quenching may be driven more by heating the gas than expelling it.

\section{Summary}

We have presented analysis based on optical spectroscopy of two samples of z$\sim$1.2 QGs; those which are compact relative to local early-type galaxies, and larger, more extended ''normal'' QGs. We find evidence that compact QGs on average have older stellar ages than normal-sized QGs. This observed density-age correlation is an empirical prediction of the progenitor bias scenario, whereby the first galaxies to evolve and quench are simply the most compact due to their early formation time, and extended QGs form later and quench later. We also study the [\ion{O}{2}] emission and X-ray properties of the two samples, finding evidence for lower incidence rate of residual energy sources in the compact sample compared to normal QGs. This is consistent with the idea that the compact sample quenched earlier, having had longer time to fade. We do not find explicit evidence for compactness-driven quenching, and suggest that future studies of galaxies at their quenching epoch may help illuminate whether the feedback or regulation of star-formation differs with stellar density.

\acknowledgements 

We thank the anonymous referee whose valuable suggestions have improved the paper significantly. CCW acknowledges support from the JWST/NIRCam contract to the University of Arizona, NAS5-02015. NC acknowledges Yale University's YCAA Prize Postdoctoral fellowship. T.L. thanks support from the National Science Foundation of China No. 11403021. This work is based on observations taken by the CANDELS Multi-cycle Treasury Program with the NASA/ESA {\it HST}. This research made use of Astropy, a community-developed
core Python package for Astronomy \citep{astropy}, and CSTACK (http://lambic.astrosen.unam.mx/cstack/) developed by Takamitsu Miyaji.

\bibliographystyle{aa}

\begin{thebibliography}{134}
\expandafter\ifx\csname natexlab\endcsname\relax\def\natexlab#1{#1}\fi

\bibitem[{{Abramson} \& {Morishita}(2016)}]{Abramson2016}
{Abramson}, L.~E. \& {Morishita}, T. 2016, ArXiv e-prints

\bibitem[{{Acquaviva} {et~al.}(2011){Acquaviva}, {Gawiser}, \&
  {Guaita}}]{Acquaviva2011}
{Acquaviva}, V., {Gawiser}, E., \& {Guaita}, L. 2011, \apj, 737, 47

\bibitem[{{Alexandroff} {et~al.}(2015){Alexandroff}, {Heckman}, {Borthakur},
  {Overzier}, \& {Leitherer}}]{Alexandroff2015}
{Alexandroff}, R.~M., {Heckman}, T.~M., {Borthakur}, S., {Overzier}, R., \&
  {Leitherer}, C. 2015, \apj, 810, 104

\bibitem[{{Astropy Collaboration} {et~al.}(2013){Astropy Collaboration},
  {Robitaille}, {Tollerud}, {Greenfield}, {Droettboom}, {Bray}, {Aldcroft},
  {Davis}, {Ginsburg}, {Price-Whelan}, {Kerzendorf}, {Conley}, {Crighton},
  {Barbary}, {Muna}, {Ferguson}, {Grollier}, {Parikh}, {Nair}, {Unther},
  {Deil}, {Woillez}, {Conseil}, {Kramer}, {Turner}, {Singer}, {Fox}, {Weaver},
  {Zabalza}, {Edwards}, {Azalee Bostroem}, {Burke}, {Casey}, {Crawford},
  {Dencheva}, {Ely}, {Jenness}, {Labrie}, {Lim}, {Pierfederici}, {Pontzen},
  {Ptak}, {Refsdal}, {Servillat}, \& {Streicher}}]{astropy}
{Astropy Collaboration}, {Robitaille}, T.~P., {Tollerud}, E.~J., {et~al.} 2013,
  \aap, 558, A33

\bibitem[{{Balestra} {et~al.}(2010){Balestra}, {Mainieri}, {Popesso},
  {Dickinson}, {Nonino}, {Rosati}, {Teimoorinia}, {Vanzella}, {Cristiani},
  {Cesarsky}, {Fosbury}, {Kuntschner}, \& {Rettura}}]{Balestra2010}
{Balestra}, I., {Mainieri}, V., {Popesso}, P., {et~al.} 2010, \aap, 512, A12

\bibitem[{{Balogh} {et~al.}(1999){Balogh}, {Morris}, {Yee}, {Carlberg}, \&
  {Ellingson}}]{Balogh1999}
{Balogh}, M.~L., {Morris}, S.~L., {Yee}, H.~K.~C., {Carlberg}, R.~G., \&
  {Ellingson}, E. 1999, \apj, 527, 54

\bibitem[{{Barro} {et~al.}(2013){Barro}, {Faber}, {P{\'e}rez-Gonz{\'a}lez},
  {Koo}, {Williams}, {Kocevski}, {Trump}, {Mozena}, {McGrath}, {van der Wel},
  {Wuyts}, {Bell}, {Croton}, {Ceverino}, {Dekel}, {Ashby}, {Cheung},
  {Ferguson}, {Fontana}, {Fang}, {Giavalisco}, {Grogin}, {Guo}, {Hathi},
  {Hopkins}, {Huang}, {Koekemoer}, {Kartaltepe}, {Lee}, {Newman}, {Porter},
  {Primack}, {Ryan}, {Rosario}, {Somerville}, {Salvato}, \& {Hsu}}]{Barro2013}
{Barro}, G., {Faber}, S.~M., {P{\'e}rez-Gonz{\'a}lez}, P.~G., {et~al.} 2013,
  \apj, 765, 104

\bibitem[{{Bell} {et~al.}(2012){Bell}, {van der Wel}, {Papovich}, {Kocevski},
  {Lotz}, {McIntosh}, {Kartaltepe}, {Faber}, {Ferguson}, {Koekemoer}, {Grogin},
  {Wuyts}, {Cheung}, {Conselice}, {Dekel}, {Dunlop}, {Giavalisco},
  {Herrington}, {Koo}, {McGrath}, {de Mello}, {Rix}, {Robaina}, \&
  {Williams}}]{Bell2012}
{Bell}, E.~F., {van der Wel}, A., {Papovich}, C., {et~al.} 2012, \apj, 753, 167

\bibitem[{{Belli} {et~al.}(2015){Belli}, {Newman}, \& {Ellis}}]{Belli2015}
{Belli}, S., {Newman}, A.~B., \& {Ellis}, R.~S. 2015, \apj, 799, 206

\bibitem[{{Bezanson} {et~al.}(2009){Bezanson}, {van Dokkum}, {Tal},
  {Marchesini}, {Kriek}, {Franx}, \& {Coppi}}]{Bezanson2009}
{Bezanson}, R., {van Dokkum}, P.~G., {Tal}, T., {et~al.} 2009, \apj, 697, 1290

\bibitem[{{Birnboim} \& {Dekel}(2003)}]{BirnboimDekel2003}
{Birnboim}, Y. \& {Dekel}, A. 2003, \mnras, 345, 349

\bibitem[{{Boroson} {et~al.}(2011){Boroson}, {Kim}, \&
  {Fabbiano}}]{Boroson2011}
{Boroson}, B., {Kim}, D.-W., \& {Fabbiano}, G. 2011, \apj, 729, 12

\bibitem[{{Bower} {et~al.}(1992){Bower}, {Lucey}, \& {Ellis}}]{Bower1992}
{Bower}, R.~G., {Lucey}, J.~R., \& {Ellis}, R.~S. 1992, \mnras, 254, 601

\bibitem[{{Bundy} {et~al.}(2006){Bundy}, {Ellis}, {Conselice}, {Taylor},
  {Cooper}, {Willmer}, {Weiner}, {Coil}, {Noeske}, \& {Eisenhardt}}]{Bundy2006}
{Bundy}, K., {Ellis}, R.~S., {Conselice}, C.~J., {et~al.} 2006, \apj, 651, 120

\bibitem[{{Cappellari} \& {Emsellem}(2004)}]{CappellariEmsellem2004}
{Cappellari}, M. \& {Emsellem}, E. 2004, \pasp, 116, 138

\bibitem[{{Cappelluti} {et~al.}(2016){Cappelluti}, {Comastri}, {Fontana},
  {Zamorani}, {Amorin}, {Castellano}, {Merlin}, {Santini}, {Elbaz},
  {Schreiber}, {Shu}, {Wang}, {Dunlop}, {Bourne}, {Bruce}, {Buitrago},
  {Micha{\l}owski}, {Derriere}, {Ferguson}, {Faber}, \&
  {Vito}}]{Cappelluti2016}
{Cappelluti}, N., {Comastri}, A., {Fontana}, A., {et~al.} 2016, \apj, 823, 95

\bibitem[{{Carollo} {et~al.}(2013){Carollo}, {Bschorr}, {Renzini}, {Lilly},
  {Capak}, {Cibinel}, {Ilbert}, {Onodera}, {Scoville}, {Cameron}, {Mobasher},
  {Sanders}, \& {Taniguchi}}]{Carollo2013}
{Carollo}, C.~M., {Bschorr}, T.~J., {Renzini}, A., {et~al.} 2013, \apj, 773,
  112

\bibitem[{{Cassata} {et~al.}(2011){Cassata}, {Giavalisco}, {Guo}, {Renzini},
  {Ferguson}, {Koekemoer}, {Salimbeni}, {Scarlata}, {Grogin}, {Conselice},
  {Dahlen}, {Lotz}, {Dickinson}, \& {Lin}}]{Cassata2011}
{Cassata}, P., {Giavalisco}, M., {Guo}, Y., {et~al.} 2011, \apj, 743, 96

\bibitem[{{Cassata} {et~al.}(2013){Cassata}, {Giavalisco}, {Williams}, {Guo},
  {Lee}, {Renzini}, {Ferguson}, {Faber}, {Barro}, {McIntosh}, {Lu}, {Bell},
  {Koo}, {Papovich}, {Ryan}, {Conselice}, {Grogin}, {Koekemoer}, \&
  {Hathi}}]{Cassata2013}
{Cassata}, P., {Giavalisco}, M., {Williams}, C.~C., {et~al.} 2013, \apj, 775,
  106

\bibitem[{{Ceverino} {et~al.}(2015){Ceverino}, {Dekel}, {Tweed}, \&
  {Primack}}]{Ceverino2015}
{Ceverino}, D., {Dekel}, A., {Tweed}, D., \& {Primack}, J. 2015, \mnras, 447,
  3291

\bibitem[{{Cimatti} {et~al.}(2008){Cimatti}, {Cassata}, {Pozzetti}, {Kurk},
  {Mignoli}, {Renzini}, {Daddi}, {Bolzonella}, {Brusa}, {Rodighiero},
  {Dickinson}, {Franceschini}, {Zamorani}, {Berta}, {Rosati}, \&
  {Halliday}}]{Cimatti2008}
{Cimatti}, A., {Cassata}, P., {Pozzetti}, L., {et~al.} 2008, \aap, 482, 21

\bibitem[{{Ciotti} \& {Ostriker}(1997)}]{CiottiOstriker1997}
{Ciotti}, L. \& {Ostriker}, J.~P. 1997, \apjl, 487, L105

\bibitem[{{Colbert} {et~al.}(2004){Colbert}, {Heckman}, {Ptak}, {Strickland},
  \& {Weaver}}]{Colbert2004}
{Colbert}, E.~J.~M., {Heckman}, T.~M., {Ptak}, A.~F., {Strickland}, D.~K., \&
  {Weaver}, K.~A. 2004, \apj, 602, 231

\bibitem[{{Daddi} {et~al.}(2005){Daddi}, {Renzini}, {Pirzkal}, {Cimatti},
  {Malhotra}, {Stiavelli}, {Xu}, {Pasquali}, {Rhoads}, {Brusa}, {di Serego
  Alighieri}, {Ferguson}, {Koekemoer}, {Moustakas}, {Panagia}, \&
  {Windhorst}}]{Daddi2005}
{Daddi}, E., {Renzini}, A., {Pirzkal}, N., {et~al.} 2005, \apj, 626, 680

\bibitem[{{Damjanov} {et~al.}(2009){Damjanov}, {McCarthy}, {Abraham},
  {Glazebrook}, {Yan}, {Mentuch}, {Le Borgne}, {Savaglio}, {Crampton},
  {Murowinski}, {Juneau}, {Carlberg}, {J{\o}rgensen}, {Roth}, {Chen}, \&
  {Marzke}}]{Damjanov2009}
{Damjanov}, I., {McCarthy}, P.~J., {Abraham}, R.~G., {et~al.} 2009, \apj, 695,
  101

\bibitem[{{Dekel} \& {Birnboim}(2006)}]{DekelBirnboim2006}
{Dekel}, A. \& {Birnboim}, Y. 2006, \mnras, 368, 2

\bibitem[{{Dekel} \& {Burkert}(2014)}]{DekelBurkert2014}
{Dekel}, A. \& {Burkert}, A. 2014, \mnras, 438, 1870

\bibitem[{{Dekel} {et~al.}(2009){Dekel}, {Sari}, \& {Ceverino}}]{Dekel2009}
{Dekel}, A., {Sari}, R., \& {Ceverino}, D. 2009, \apj, 703, 785

\bibitem[{{Diamond-Stanic} {et~al.}(2012){Diamond-Stanic}, {Moustakas},
  {Tremonti}, {Coil}, {Hickox}, {Robaina}, {Rudnick}, \&
  {Sell}}]{DiamondStanic2012}
{Diamond-Stanic}, A.~M., {Moustakas}, J., {Tremonti}, C.~A., {et~al.} 2012,
  \apjl, 755, L26

\bibitem[{{Dressler} {et~al.}(2004){Dressler}, {Oemler}, {Poggianti}, {Smail},
  {Trager}, {Shectman}, {Couch}, \& {Ellis}}]{Dressler2004}
{Dressler}, A., {Oemler}, Jr., A., {Poggianti}, B.~M., {et~al.} 2004, \apj,
  617, 867

\bibitem[{{Fabbiano} {et~al.}(1992){Fabbiano}, {Kim}, \&
  {Trinchieri}}]{Fabbiano1992}
{Fabbiano}, G., {Kim}, D.-W., \& {Trinchieri}, G. 1992, \apjs, 80, 531

\bibitem[{{Fabian}(2012)}]{Fabian2012}
{Fabian}, A.~C. 2012, \araa, 50, 455

\bibitem[{{Fagioli} {et~al.}(2016){Fagioli}, {Carollo}, {Renzini}, {Lilly},
  {Onodera}, \& {Tacchella}}]{Fagioli2016}
{Fagioli}, M., {Carollo}, C.~M., {Renzini}, A., {et~al.} 2016, ArXiv e-prints

\bibitem[{{Fontana} {et~al.}(2009){Fontana}, {Santini}, {Grazian},
  {Pentericci}, {Fiore}, {Castellano}, {Giallongo}, {Menci}, {Salimbeni},
  {Cristiani}, {Nonino}, \& {Vanzella}}]{Fontana2009}
{Fontana}, A., {Santini}, P., {Grazian}, A., {et~al.} 2009, \aap, 501, 15

\bibitem[{{Fragos} {et~al.}(2013){Fragos}, {Lehmer}, {Tremmel}, {Tzanavaris},
  {Basu-Zych}, {Belczynski}, {Hornschemeier}, {Jenkins}, {Kalogera}, {Ptak}, \&
  {Zezas}}]{Fragos2013}
{Fragos}, T., {Lehmer}, B., {Tremmel}, M., {et~al.} 2013, \apj, 764, 41

\bibitem[{{Franx} {et~al.}(2008){Franx}, {van Dokkum}, {Schreiber}, {Wuyts},
  {Labb{\'e}}, \& {Toft}}]{Franx2008}
{Franx}, M., {van Dokkum}, P.~G., {Schreiber}, N.~M.~F., {et~al.} 2008, \apj,
  688, 770

\bibitem[{{Gallazzi} {et~al.}(2014){Gallazzi}, {Bell}, {Zibetti}, {Brinchmann},
  \& {Kelson}}]{Gallazzi2014}
{Gallazzi}, A., {Bell}, E.~F., {Zibetti}, S., {Brinchmann}, J., \& {Kelson},
  D.~D. 2014, \apj, 788, 72

\bibitem[{{Gallazzi} {et~al.}(2005){Gallazzi}, {Charlot}, {Brinchmann},
  {White}, \& {Tremonti}}]{Gallazzi2005}
{Gallazzi}, A., {Charlot}, S., {Brinchmann}, J., {White}, S.~D.~M., \&
  {Tremonti}, C.~A. 2005, \mnras, 362, 41

\bibitem[{{Giavalisco} {et~al.}(2004){Giavalisco}, {Ferguson}, {Koekemoer},
  {Dickinson}, {Alexander}, {Bauer}, {Bergeron}, {Biagetti}, {Brandt},
  {Casertano}, {Cesarsky}, {Chatzichristou}, {Conselice}, {Cristiani}, {Da
  Costa}, {Dahlen}, {de Mello}, {Eisenhardt}, {Erben}, {Fall}, {Fassnacht},
  {Fosbury}, {Fruchter}, {Gardner}, {Grogin}, {Hook}, {Hornschemeier}, {Idzi},
  {Jogee}, {Kretchmer}, {Laidler}, {Lee}, {Livio}, {Lucas}, {Madau},
  {Mobasher}, {Moustakas}, {Nonino}, {Padovani}, {Papovich}, {Park},
  {Ravindranath}, {Renzini}, {Richardson}, {Riess}, {Rosati}, {Schirmer},
  {Schreier}, {Somerville}, {Spinrad}, {Stern}, {Stiavelli}, {Strolger},
  {Urry}, {Vandame}, {Williams}, \& {Wolf}}]{Giavalisco2004}
{Giavalisco}, M., {Ferguson}, H.~C., {Koekemoer}, A.~M., {et~al.} 2004, \apjl,
  600, L93

\bibitem[{{Gilfanov}(2004)}]{Gilfanov2004}
{Gilfanov}, M. 2004, \mnras, 349, 146

\bibitem[{{Girardi} {et~al.}(2000){Girardi}, {Bressan}, {Bertelli}, \&
  {Chiosi}}]{Girardi2000}
{Girardi}, L., {Bressan}, A., {Bertelli}, G., \& {Chiosi}, C. 2000, \aaps, 141,
  371

\bibitem[{{Gobat} {et~al.}(2012){Gobat}, {Strazzullo}, {Daddi}, {Onodera},
  {Renzini}, {B{\'e}thermin}, {Dickinson}, {Carollo}, \& {Cimatti}}]{Gobat2012}
{Gobat}, R., {Strazzullo}, V., {Daddi}, E., {et~al.} 2012, \apjl, 759, L44

\bibitem[{{Goulding} {et~al.}(2016){Goulding}, {Greene}, {Ma}, {Veale},
  {Bogdan}, {Nyland}, {Blakeslee}, {McConnell}, \& {Thomas}}]{Goulding2016}
{Goulding}, A.~D., {Greene}, J.~E., {Ma}, C.-P., {et~al.} 2016, \apj, 826, 167

\bibitem[{{Granato} {et~al.}(2004){Granato}, {De Zotti}, {Silva}, {Bressan}, \&
  {Danese}}]{Granato2004}
{Granato}, G.~L., {De Zotti}, G., {Silva}, L., {Bressan}, A., \& {Danese}, L.
  2004, \apj, 600, 580

\bibitem[{{Grogin} {et~al.}(2011){Grogin}, {Kocevski}, {Faber}, {Ferguson},
  {Koekemoer}, {Riess}, {Acquaviva}, {Alexander}, {Almaini}, {Ashby}, {Barden},
  {Bell}, {Bournaud}, {Brown}, {Caputi}, {Casertano}, {Cassata}, {Castellano},
  {Challis}, {Chary}, {Cheung}, {Cirasuolo}, {Conselice}, {Roshan Cooray},
  {Croton}, {Daddi}, {Dahlen}, {Dav{\'e}}, {de Mello}, {Dekel}, {Dickinson},
  {Dolch}, {Donley}, {Dunlop}, {Dutton}, {Elbaz}, {Fazio}, {Filippenko},
  {Finkelstein}, {Fontana}, {Gardner}, {Garnavich}, {Gawiser}, {Giavalisco},
  {Grazian}, {Guo}, {Hathi}, {H{\"a}ussler}, {Hopkins}, {Huang}, {Huang},
  {Jha}, {Kartaltepe}, {Kirshner}, {Koo}, {Lai}, {Lee}, {Li}, {Lotz}, {Lucas},
  {Madau}, {McCarthy}, {McGrath}, {McIntosh}, {McLure}, {Mobasher},
  {Moustakas}, {Mozena}, {Nandra}, {Newman}, {Niemi}, {Noeske}, {Papovich},
  {Pentericci}, {Pope}, {Primack}, {Rajan}, {Ravindranath}, {Reddy}, {Renzini},
  {Rix}, {Robaina}, {Rodney}, {Rosario}, {Rosati}, {Salimbeni}, {Scarlata},
  {Siana}, {Simard}, {Smidt}, {Somerville}, {Spinrad}, {Straughn}, {Strolger},
  {Telford}, {Teplitz}, {Trump}, {van der Wel}, {Villforth}, {Wechsler},
  {Weiner}, {Wiklind}, {Wild}, {Wilson}, {Wuyts}, {Yan}, \& {Yun}}]{Grogin2011}
{Grogin}, N.~A., {Kocevski}, D.~D., {Faber}, S.~M., {et~al.} 2011, \apjs, 197,
  35

\bibitem[{{Guo} {et~al.}(2013){Guo}, {Ferguson}, {Giavalisco}, {Barro},
  {Willner}, {Ashby}, {Dahlen}, {Donley}, {Faber}, {Fontana}, {Galametz},
  {Grazian}, {Huang}, {Kocevski}, {Koekemoer}, {Koo}, {McGrath}, {Peth},
  {Salvato}, {Wuyts}, {Castellano}, {Cooray}, {Dickinson}, {Dunlop}, {Fazio},
  {Gardner}, {Gawiser}, {Grogin}, {Hathi}, {Hsu}, {Lee}, {Lucas}, {Mobasher},
  {Nandra}, {Newman}, \& {van der Wel}}]{Guo2013}
{Guo}, Y., {Ferguson}, H.~C., {Giavalisco}, M., {et~al.} 2013, \apjs, 207, 24

\bibitem[{{Guo} {et~al.}(2012){Guo}, {Giavalisco}, {Cassata}, {Ferguson},
  {Williams}, {Dickinson}, {Koekemoer}, {Grogin}, {Chary}, {Messias}, {Tundo},
  {Lin}, {Lee}, {Salimbeni}, {Fontana}, {Grazian}, {Kocevski}, {Lee},
  {Villanueva}, \& {van der Wel}}]{Guo2012}
{Guo}, Y., {Giavalisco}, M., {Cassata}, P., {et~al.} 2012, \apj, 749, 149

\bibitem[{{Hasinger}(2008)}]{Hasinger2008}
{Hasinger}, G. 2008, \aap, 490, 905

\bibitem[{{Heavens} {et~al.}(2004){Heavens}, {Panter}, {Jimenez}, \&
  {Dunlop}}]{Heavens2004}
{Heavens}, A., {Panter}, B., {Jimenez}, R., \& {Dunlop}, J. 2004, \nat, 428,
  625

\bibitem[{{Heckman}(1980)}]{Heckman1980}
{Heckman}, T.~M. 1980, \aap, 87, 152

\bibitem[{{Heckman} \& {Borthakur}(2016)}]{HeckmanBorthakur2016}
{Heckman}, T.~M. \& {Borthakur}, S. 2016, \apj, 822, 9

\bibitem[{{Hopkins} {et~al.}(2010){Hopkins}, {Murray}, {Quataert}, \&
  {Thompson}}]{Hopkins2010}
{Hopkins}, P.~F., {Murray}, N., {Quataert}, E., \& {Thompson}, T.~A. 2010,
  \mnras, 401, L19

\bibitem[{{Jacoby} {et~al.}(1984){Jacoby}, {Hunter}, \&
  {Christian}}]{Jacoby1984}
{Jacoby}, G.~H., {Hunter}, D.~A., \& {Christian}, C.~A. 1984, \apjs, 56, 257

\bibitem[{{Johansson} {et~al.}(2012){Johansson}, {Naab}, \&
  {Ostriker}}]{Johansson2012}
{Johansson}, P.~H., {Naab}, T., \& {Ostriker}, J.~P. 2012, \apj, 754, 115

\bibitem[{{Kauffmann} {et~al.}(2003{\natexlab{a}}){Kauffmann}, {Heckman},
  {White}, {Charlot}, {Tremonti}, {Brinchmann}, {Bruzual}, {Peng}, {Seibert},
  {Bernardi}, {Blanton}, {Brinkmann}, {Castander}, {Cs{\'a}bai}, {Fukugita},
  {Ivezic}, {Munn}, {Nichol}, {Padmanabhan}, {Thakar}, {Weinberg}, \&
  {York}}]{Kauffmann2003}
{Kauffmann}, G., {Heckman}, T.~M., {White}, S.~D.~M., {et~al.}
  2003{\natexlab{a}}, \mnras, 341, 33

\bibitem[{{Kauffmann} {et~al.}(2003{\natexlab{b}}){Kauffmann}, {Heckman},
  {White}, {Charlot}, {Tremonti}, {Peng}, {Seibert}, {Brinkmann}, {Nichol},
  {SubbaRao}, \& {York}}]{Kauffmann2003b}
{Kauffmann}, G., {Heckman}, T.~M., {White}, S.~D.~M., {et~al.}
  2003{\natexlab{b}}, \mnras, 341, 54

\bibitem[{{Keating} {et~al.}(2015){Keating}, {Abraham}, {Schiavon}, {Graves},
  {Damjanov}, {Yan}, {Newman}, \& {Simard}}]{Keating2015}
{Keating}, S.~K., {Abraham}, R.~G., {Schiavon}, R., {et~al.} 2015, \apj, 798,
  26

\bibitem[{{Kennicutt}(1998)}]{Kennicutt1998}
{Kennicutt}, Jr., R.~C. 1998, \araa, 36, 189

\bibitem[{{Kewley} {et~al.}(2004){Kewley}, {Geller}, \& {Jansen}}]{Kewley2004}
{Kewley}, L.~J., {Geller}, M.~J., \& {Jansen}, R.~A. 2004, \aj, 127, 2002

\bibitem[{{Kim} \& {Fabbiano}(2004)}]{KimFabbiano2004}
{Kim}, D.-W. \& {Fabbiano}, G. 2004, \apj, 611, 846

\bibitem[{{Kim} \& {Fabbiano}(2013)}]{KimFabbiano2013}
{Kim}, D.-W. \& {Fabbiano}, G. 2013, \apj, 776, 116

\bibitem[{{Koekemoer} {et~al.}(2011){Koekemoer}, {Faber}, {Ferguson}, {Grogin},
  {Kocevski}, {Koo}, {Lai}, {Lotz}, {Lucas}, {McGrath}, {Ogaz}, {Rajan},
  {Riess}, {Rodney}, {Strolger}, {Casertano}, {Castellano}, {Dahlen},
  {Dickinson}, {Dolch}, {Fontana}, {Giavalisco}, {Grazian}, {Guo}, {Hathi},
  {Huang}, {van der Wel}, {Yan}, {Acquaviva}, {Alexander}, {Almaini}, {Ashby},
  {Barden}, {Bell}, {Bournaud}, {Brown}, {Caputi}, {Cassata}, {Challis},
  {Chary}, {Cheung}, {Cirasuolo}, {Conselice}, {Roshan Cooray}, {Croton},
  {Daddi}, {Dav{\'e}}, {de Mello}, {de Ravel}, {Dekel}, {Donley}, {Dunlop},
  {Dutton}, {Elbaz}, {Fazio}, {Filippenko}, {Finkelstein}, {Frazer}, {Gardner},
  {Garnavich}, {Gawiser}, {Gruetzbauch}, {Hartley}, {H{\"a}ussler},
  {Herrington}, {Hopkins}, {Huang}, {Jha}, {Johnson}, {Kartaltepe},
  {Khostovan}, {Kirshner}, {Lani}, {Lee}, {Li}, {Madau}, {McCarthy},
  {McIntosh}, {McLure}, {McPartland}, {Mobasher}, {Moreira}, {Mortlock},
  {Moustakas}, {Mozena}, {Nandra}, {Newman}, {Nielsen}, {Niemi}, {Noeske},
  {Papovich}, {Pentericci}, {Pope}, {Primack}, {Ravindranath}, {Reddy},
  {Renzini}, {Rix}, {Robaina}, {Rosario}, {Rosati}, {Salimbeni}, {Scarlata},
  {Siana}, {Simard}, {Smidt}, {Snyder}, {Somerville}, {Spinrad}, {Straughn},
  {Telford}, {Teplitz}, {Trump}, {Vargas}, {Villforth}, {Wagner}, {Wandro},
  {Wechsler}, {Weiner}, {Wiklind}, {Wild}, {Wilson}, {Wuyts}, \&
  {Yun}}]{Koekemoer2011}
{Koekemoer}, A.~M., {Faber}, S.~M., {Ferguson}, H.~C., {et~al.} 2011, \apjs,
  197, 36

\bibitem[{{Kurk} {et~al.}(2009){Kurk}, {Cimatti}, {Daddi}, {Mignoli},
  {Bolzonella}, {Pozzetti}, {Cassata}, {Halliday}, {Zamorani}, {Berta},
  {Brusa}, {Dickinson}, {Franceschini}, {Rodighiero}, {Rosati}, \&
  {Renzini}}]{Kurk2009}
{Kurk}, J., {Cimatti}, A., {Daddi}, E., {et~al.} 2009, The Messenger, 135, 40

\bibitem[{{Kurk} {et~al.}(2013){Kurk}, {Cimatti}, {Daddi}, {Mignoli},
  {Pozzetti}, {Dickinson}, {Bolzonella}, {Zamorani}, {Cassata}, {Rodighiero},
  {Franceschini}, {Renzini}, {Rosati}, {Halliday}, \& {Berta}}]{Kurk2013}
{Kurk}, J., {Cimatti}, A., {Daddi}, E., {et~al.} 2013, \aap, 549, A63

\bibitem[{{Le Borgne} {et~al.}(2003){Le Borgne}, {Bruzual}, {Pell{\'o}},
  {Lan{\c c}on}, {Rocca-Volmerange}, {Sanahuja}, {Schaerer}, {Soubiran}, \&
  {V{\'{\i}}lchez-G{\'o}mez}}]{LeBorgne2003}
{Le Borgne}, J.-F., {Bruzual}, G., {Pell{\'o}}, R., {et~al.} 2003, \aap, 402,
  433

\bibitem[{{Lehmer} {et~al.}(2007){Lehmer}, {Brandt}, {Alexander}, {Bell},
  {McIntosh}, {Bauer}, {Hasinger}, {Mainieri}, {Miyaji}, {Schneider}, \&
  {Steffen}}]{Lehmer2007}
{Lehmer}, B.~D., {Brandt}, W.~N., {Alexander}, D.~M., {et~al.} 2007, \apj, 657,
  681

\bibitem[{{Lilly} \& {Carollo}(2016)}]{LillyCarollo2016}
{Lilly}, S.~J. \& {Carollo}, C.~M. 2016, ArXiv e-prints

\bibitem[{{L{\'o}pez-Sanjuan} {et~al.}(2012){L{\'o}pez-Sanjuan}, {Le
  F{\`e}vre}, {Ilbert}, {Tasca}, {Bridge}, {Cucciati}, {Kampczyk}, {Pozzetti},
  {Xu}, {Carollo}, {Contini}, {Kneib}, {Lilly}, {Mainieri}, {Renzini},
  {Sanders}, {Scodeggio}, {Scoville}, {Taniguchi}, {Zamorani}, {Aussel},
  {Bardelli}, {Bolzonella}, {Bongiorno}, {Capak}, {Caputi}, {de la Torre}, {de
  Ravel}, {Franzetti}, {Garilli}, {Iovino}, {Knobel}, {Kova{\v c}},
  {Lamareille}, {Le Borgne}, {Le Brun}, {Le Floc'h}, {Maier}, {McCracken},
  {Mignoli}, {Pell{\'o}}, {Peng}, {P{\'e}rez-Montero}, {Presotto},
  {Ricciardelli}, {Salvato}, {Silverman}, {Tanaka}, {Tresse}, {Vergani},
  {Zucca}, {Barnes}, {Bordoloi}, {Cappi}, {Cimatti}, {Coppa}, {Koekemoer},
  {Liu}, {Moresco}, {Nair}, {Oesch}, {Schawinski}, \&
  {Welikala}}]{LopezSanjuan2012}
{L{\'o}pez-Sanjuan}, C., {Le F{\`e}vre}, O., {Ilbert}, O., {et~al.} 2012, \aap,
  548, A7

\bibitem[{{Luo} {et~al.}(2016){Luo}, {Brandt}, {Xue}, {Lehmer}, {Alexander},
  {Bauer}, {Vito}, {Yang}, {Basu-Zych}, {Comastri}, {Gilli}, {Gu},
  {Hornschemeier}, {Koekemoer}, {Liu}, {Mainieri}, {Paolillo}, {Ranalli},
  {Rosati}, {Schneider}, {Shemmer}, {Smail}, {Sun}, {Tozzi}, {Vignali}, \&
  {Wang}}]{Luo2016}
{Luo}, B., {Brandt}, W.~N., {Xue}, Y.~Q., {et~al.} 2016, ArXiv e-prints

\bibitem[{{Mainieri} {et~al.}(2002){Mainieri}, {Bergeron}, {Hasinger},
  {Lehmann}, {Rosati}, {Schmidt}, {Szokoly}, \& {Della Ceca}}]{Mainieri2002}
{Mainieri}, V., {Bergeron}, J., {Hasinger}, G., {et~al.} 2002, \aap, 393, 425

\bibitem[{{Matsumoto} {et~al.}(1997){Matsumoto}, {Koyama}, {Awaki}, {Tsuru},
  {Loewenstein}, \& {Matsushita}}]{Matsumoto1997}
{Matsumoto}, H., {Koyama}, K., {Awaki}, H., {et~al.} 1997, \apj, 482, 133

\bibitem[{{Mo} {et~al.}(1998){Mo}, {Mao}, \& {White}}]{Mo1998}
{Mo}, H.~J., {Mao}, S., \& {White}, S.~D.~M. 1998, \mnras, 295, 319

\bibitem[{{Muzzin} {et~al.}(2013){Muzzin}, {Marchesini}, {Stefanon}, {Franx},
  {McCracken}, {Milvang-Jensen}, {Dunlop}, {Fynbo}, {Brammer}, {Labbe}, \& {van
  Dokkum}}]{Muzzin2013}
{Muzzin}, A., {Marchesini}, D., {Stefanon}, M., {et~al.} 2013, ArXiv e-prints

\bibitem[{{Nelson} {et~al.}(2014){Nelson}, {van Dokkum}, {Franx}, {Brammer},
  {Momcheva}, {Schreiber}, {da Cunha}, {Tacconi}, {Bezanson}, {Kirkpatrick},
  {Leja}, {Rix}, {Skelton}, {van der Wel}, {Whitaker}, \& {Wuyts}}]{Nelson2014}
{Nelson}, E., {van Dokkum}, P., {Franx}, M., {et~al.} 2014, \nat, 513, 394

\bibitem[{{Omand} {et~al.}(2014){Omand}, {Balogh}, \& {Poggianti}}]{Omand2014}
{Omand}, C.~M.~B., {Balogh}, M.~L., \& {Poggianti}, B.~M. 2014, \mnras, 440,
  843

\bibitem[{{Patel} {et~al.}(2013){Patel}, {van Dokkum}, {Franx}, {Quadri},
  {Muzzin}, {Marchesini}, {Williams}, {Holden}, \& {Stefanon}}]{Patel2013}
{Patel}, S.~G., {van Dokkum}, P.~G., {Franx}, M., {et~al.} 2013, \apj, 766, 15

\bibitem[{{Peng} {et~al.}(2002){Peng}, {Ho}, {Impey}, \& {Rix}}]{Peng2002}
{Peng}, C.~Y., {Ho}, L.~C., {Impey}, C.~D., \& {Rix}, H.-W. 2002, \aj, 124, 266

\bibitem[{{Peng} {et~al.}(2010){Peng}, {Lilly}, {Kova{\v c}}, {Bolzonella},
  {Pozzetti}, {Renzini}, {Zamorani}, {Ilbert}, {Knobel}, {Iovino}, {Maier},
  {Cucciati}, {Tasca}, {Carollo}, {Silverman}, {Kampczyk}, {de Ravel},
  {Sanders}, {Scoville}, {Contini}, {Mainieri}, {Scodeggio}, {Kneib}, {Le
  F{\`e}vre}, {Bardelli}, {Bongiorno}, {Caputi}, {Coppa}, {de la Torre},
  {Franzetti}, {Garilli}, {Lamareille}, {Le Borgne}, {Le Brun}, {Mignoli},
  {Perez Montero}, {Pello}, {Ricciardelli}, {Tanaka}, {Tresse}, {Vergani},
  {Welikala}, {Zucca}, {Oesch}, {Abbas}, {Barnes}, {Bordoloi}, {Bottini},
  {Cappi}, {Cassata}, {Cimatti}, {Fumana}, {Hasinger}, {Koekemoer},
  {Leauthaud}, {Maccagni}, {Marinoni}, {McCracken}, {Memeo}, {Meneux}, {Nair},
  {Porciani}, {Presotto}, \& {Scaramella}}]{Peng2010}
{Peng}, Y.-j., {Lilly}, S.~J., {Kova{\v c}}, K., {et~al.} 2010, \apj, 721, 193

\bibitem[{{Pickles}(1998)}]{Pickles1998}
{Pickles}, A.~J. 1998, \pasp, 110, 863

\bibitem[{{Poggianti} {et~al.}(2013){Poggianti}, {Calvi}, {Bindoni},
  {D'Onofrio}, {Moretti}, {Valentinuzzi}, {Fasano}, {Fritz}, {De Lucia},
  {Vulcani}, {Bettoni}, {Gullieuszik}, \& {Omizzolo}}]{Poggianti2013}
{Poggianti}, B.~M., {Calvi}, R., {Bindoni}, D., {et~al.} 2013, \apj, 762, 77

\bibitem[{{Popesso} {et~al.}(2009){Popesso}, {Dickinson}, {Nonino}, {Vanzella},
  {Daddi}, {Fosbury}, {Kuntschner}, {Mainieri}, {Cristiani}, {Cesarsky},
  {Giavalisco}, {Renzini}, \& {GOODS Team}}]{Popesso2009}
{Popesso}, P., {Dickinson}, M., {Nonino}, M., {et~al.} 2009, \aap, 494, 443

\bibitem[{{Prochaska} {et~al.}(2007){Prochaska}, {Rose}, {Caldwell},
  {Castilho}, {Concannon}, {Harding}, {Morrison}, \&
  {Schiavon}}]{Prochaska2007}
{Prochaska}, L.~C., {Rose}, J.~A., {Caldwell}, N., {et~al.} 2007, \aj, 134, 401

\bibitem[{{Renzini}(2006)}]{Renzini2006}
{Renzini}, A. 2006, \araa, 44, 141

\bibitem[{{Renzini} {et~al.}(1993){Renzini}, {Ciotti}, {D'Ercole}, \&
  {Pellegrini}}]{Renzini1993}
{Renzini}, A., {Ciotti}, L., {D'Ercole}, A., \& {Pellegrini}, S. 1993, \apj,
  419, 52

\bibitem[{{Salpeter}(1955)}]{Salpeter1955}
{Salpeter}, E.~E. 1955, \apj, 121, 161

\bibitem[{{Saracco} {et~al.}(2009){Saracco}, {Longhetti}, \&
  {Andreon}}]{Saracco2009}
{Saracco}, P., {Longhetti}, M., \& {Andreon}, S. 2009, \mnras, 392, 718

\bibitem[{{Saracco} {et~al.}(2011){Saracco}, {Longhetti}, \&
  {Gargiulo}}]{Saracco2011}
{Saracco}, P., {Longhetti}, M., \& {Gargiulo}, A. 2011, \mnras, 412, 2707

\bibitem[{{Schiavon} {et~al.}(2002){Schiavon}, {Faber}, {Castilho}, \&
  {Rose}}]{Schiavon2002a}
{Schiavon}, R.~P., {Faber}, S.~M., {Castilho}, B.~V., \& {Rose}, J.~A. 2002,
  \apj, 580, 850

\bibitem[{{Sell} {et~al.}(2014){Sell}, {Tremonti}, {Hickox}, {Diamond-Stanic},
  {Moustakas}, {Coil}, {Williams}, {Rudnick}, {Robaina}, {Geach}, {Heinz}, \&
  {Wilcots}}]{Sell2014}
{Sell}, P.~H., {Tremonti}, C.~A., {Hickox}, R.~C., {et~al.} 2014, \mnras, 441,
  3417

\bibitem[{{Shen} {et~al.}(2003){Shen}, {Mo}, {White}, {Blanton}, {Kauffmann},
  {Voges}, {Brinkmann}, \& {Csabai}}]{Shen2003}
{Shen}, S., {Mo}, H.~J., {White}, S.~D.~M., {et~al.} 2003, \mnras, 343, 978

\bibitem[{{Singh} {et~al.}(2013){Singh}, {van de Ven}, {Jahnke}, {Lyubenova},
  {Falc{\'o}n-Barroso}, {Alves}, {Cid Fernandes}, {Galbany},
  {Garc{\'{\i}}a-Benito}, {Husemann}, {Kennicutt}, {Marino}, {M{\'a}rquez},
  {Masegosa}, {Mast}, {Pasquali}, {S{\'a}nchez}, {Walcher}, {Wild}, {Wisotzki},
  \& {Ziegler}}]{Singh2013}
{Singh}, R., {van de Ven}, G., {Jahnke}, K., {et~al.} 2013, \aap, 558, A43

\bibitem[{{Sivakoff} {et~al.}(2004){Sivakoff}, {Sarazin}, \&
  {Carlin}}]{Sivakoff2004}
{Sivakoff}, G.~R., {Sarazin}, C.~L., \& {Carlin}, J.~L. 2004, \apj, 617, 262

\bibitem[{{Spilker} {et~al.}(2016){Spilker}, {Bezanson}, {Marrone}, {Weiner},
  {Whitaker}, \& {Williams}}]{Spilker2016}
{Spilker}, J.~S., {Bezanson}, R., {Marrone}, D.~P., {et~al.} 2016, ArXiv
  e-prints

\bibitem[{{Stefanon} {et~al.}(2013){Stefanon}, {Marchesini}, {Rudnick},
  {Brammer}, \& {Whitaker}}]{Stefanon2013}
{Stefanon}, M., {Marchesini}, D., {Rudnick}, G.~H., {Brammer}, G.~B., \&
  {Whitaker}, K.~E. 2013, \apj, 768, 92

\bibitem[{{Straatman} {et~al.}(2014){Straatman}, {Labb{\'e}}, {Spitler},
  {Allen}, {Altieri}, {Brammer}, {Dickinson}, {van Dokkum}, {Inami},
  {Glazebrook}, {Kacprzak}, {Kawinwanichakij}, {Kelson}, {McCarthy},
  {Mehrtens}, {Monson}, {Murphy}, {Papovich}, {Persson}, {Quadri}, {Rees},
  {Tomczak}, {Tran}, \& {Tilvi}}]{Straatman2014}
{Straatman}, C.~M.~S., {Labb{\'e}}, I., {Spitler}, L.~R., {et~al.} 2014, \apjl,
  783, L14

\bibitem[{{Strateva} {et~al.}(2001){Strateva}, {Ivezi{\'c}}, {Knapp},
  {Narayanan}, {Strauss}, {Gunn}, {Lupton}, {Schlegel}, {Bahcall}, {Brinkmann},
  {Brunner}, {Budav{\'a}ri}, {Csabai}, {Castander}, {Doi}, {Fukugita}, {Gy{\H
  o}ry}, {Hamabe}, {Hennessy}, {Ichikawa}, {Kunszt}, {Lamb}, {McKay},
  {Okamura}, {Racusin}, {Sekiguchi}, {Schneider}, {Shimasaku}, \&
  {York}}]{Strateva2001}
{Strateva}, I., {Ivezi{\'c}}, {\v Z}., {Knapp}, G.~R., {et~al.} 2001, \aj, 122,
  1861

\bibitem[{{Szokoly} {et~al.}(2004){Szokoly}, {Bergeron}, {Hasinger}, {Lehmann},
  {Kewley}, {Mainieri}, {Nonino}, {Rosati}, {Giacconi}, {Gilli}, {Gilmozzi},
  {Norman}, {Romaniello}, {Schreier}, {Tozzi}, {Wang}, {Zheng}, \&
  {Zirm}}]{Szokoly2004}
{Szokoly}, G.~P., {Bergeron}, J., {Hasinger}, G., {et~al.} 2004, \apjs, 155,
  271

\bibitem[{{Taniguchi} {et~al.}(2015){Taniguchi}, {Kajisawa}, {Kobayashi},
  {Nagao}, {Shioya}, {Scoville}, {Sanders}, {Capak}, {Koekemoer}, {Toft},
  {McCracken}, {Le F{\`e}vre}, {Tasca}, {Sheth}, {Renzini}, {Lilly}, {Carollo},
  {Kova{\v c}}, {Ilbert}, {Schinnerer}, {Fu}, {Tresse}, {Griffiths}, \&
  {Civano}}]{Taniguchi2015}
{Taniguchi}, Y., {Kajisawa}, M., {Kobayashi}, M.~A.~R., {et~al.} 2015, \apjl,
  809, L7

\bibitem[{{Teimoorinia} {et~al.}(2016){Teimoorinia}, {Bluck}, \&
  {Ellison}}]{Teimoorinia2016}
{Teimoorinia}, H., {Bluck}, A.~F.~L., \& {Ellison}, S.~L. 2016, \mnras, 457,
  2086

\bibitem[{{Thomas} {et~al.}(2005){Thomas}, {Maraston}, {Bender}, \& {Mendes de
  Oliveira}}]{Thomas2005}
{Thomas}, D., {Maraston}, C., {Bender}, R., \& {Mendes de Oliveira}, C. 2005,
  \apj, 621, 673

\bibitem[{{Thomas} {et~al.}(2010){Thomas}, {Maraston}, {Schawinski}, {Sarzi},
  \& {Silk}}]{Thomas2010}
{Thomas}, D., {Maraston}, C., {Schawinski}, K., {Sarzi}, M., \& {Silk}, J.
  2010, \mnras, 404, 1775

\bibitem[{{Toft} {et~al.}(2007){Toft}, {van Dokkum}, {Franx}, {Labbe},
  {F{\"o}rster Schreiber}, {Wuyts}, {Webb}, {Rudnick}, {Zirm}, {Kriek}, {van
  der Werf}, {Blakeslee}, {Illingworth}, {Rix}, {Papovich}, \&
  {Moorwood}}]{Toft2007}
{Toft}, S., {van Dokkum}, P., {Franx}, M., {et~al.} 2007, \apj, 671, 285

\bibitem[{{Treu} {et~al.}(2002){Treu}, {Stiavelli}, {Casertano}, {M{\o}ller},
  \& {Bertin}}]{Treu2002}
{Treu}, T., {Stiavelli}, M., {Casertano}, S., {M{\o}ller}, P., \& {Bertin}, G.
  2002, \apjl, 564, L13

\bibitem[{{Trujillo} {et~al.}(2011){Trujillo}, {Ferreras}, \& {de La
  Rosa}}]{Trujillo2011}
{Trujillo}, I., {Ferreras}, I., \& {de La Rosa}, I.~G. 2011, \mnras, 415, 3903

\bibitem[{{Trujillo} {et~al.}(2006){Trujillo}, {Feulner}, {Goranova}, {Hopp},
  {Longhetti}, {Saracco}, {Bender}, {Braito}, {Della Ceca}, {Drory},
  {Mannucci}, \& {Severgnini}}]{Trujillo2006}
{Trujillo}, I., {Feulner}, G., {Goranova}, Y., {et~al.} 2006, \mnras, 373, L36

\bibitem[{{Valentinuzzi} {et~al.}(2010{\natexlab{a}}){Valentinuzzi}, {Fritz},
  {Poggianti}, {Cava}, {Bettoni}, {Fasano}, {D'Onofrio}, {Couch}, {Dressler},
  {Moles}, {Moretti}, {Omizzolo}, {Kj{\ae}rgaard}, {Vanzella}, \&
  {Varela}}]{Valentinuzzi2010a}
{Valentinuzzi}, T., {Fritz}, J., {Poggianti}, B.~M., {et~al.}
  2010{\natexlab{a}}, \apj, 712, 226

\bibitem[{{Valentinuzzi} {et~al.}(2010{\natexlab{b}}){Valentinuzzi},
  {Poggianti}, {Saglia}, {Arag{\'o}n-Salamanca}, {Simard},
  {S{\'a}nchez-Bl{\'a}zquez}, {D'onofrio}, {Cava}, {Couch}, {Fritz}, {Moretti},
  \& {Vulcani}}]{Valentinuzzi2010b}
{Valentinuzzi}, T., {Poggianti}, B.~M., {Saglia}, R.~P., {et~al.}
  2010{\natexlab{b}}, \apjl, 721, L19

\bibitem[{{van de Voort} {et~al.}(2016){van de Voort}, {Quataert}, {Hopkins},
  {Faucher-Gigu{\`e}re}, {Feldmann}, {Kere{\v s}}, {Chan}, \&
  {Hafen}}]{vandeVoort2016}
{van de Voort}, F., {Quataert}, E., {Hopkins}, P.~F., {et~al.} 2016, \mnras,
  463, 4533

\bibitem[{{van der Wel} {et~al.}(2012){van der Wel}, {Bell}, {H{\"a}ussler},
  {McGrath}, {Chang}, {Guo}, {McIntosh}, {Rix}, {Barden}, {Cheung}, {Faber},
  {Ferguson}, {Galametz}, {Grogin}, {Hartley}, {Kartaltepe}, {Kocevski},
  {Koekemoer}, {Lotz}, {Mozena}, {Peth}, \& {Peng}}]{vanderWel2012}
{van der Wel}, A., {Bell}, E.~F., {H{\"a}ussler}, B., {et~al.} 2012, \apjs,
  203, 24

\bibitem[{{van der Wel} {et~al.}(2014){van der Wel}, {Franx}, {van Dokkum},
  {Skelton}, {Momcheva}, {Whitaker}, {Brammer}, {Bell}, {Rix}, {Wuyts},
  {Ferguson}, {Holden}, {Barro}, {Koekemoer}, {Chang}, {McGrath},
  {H{\"a}ussler}, {Dekel}, {Behroozi}, {Fumagalli}, {Leja}, {Lundgren},
  {Maseda}, {Nelson}, {Wake}, {Patel}, {Labb{\'e}}, {Faber}, {Grogin}, \&
  {Kocevski}}]{vanderWel2014}
{van der Wel}, A., {Franx}, M., {van Dokkum}, P.~G., {et~al.} 2014, \apj, 788,
  28

\bibitem[{{van der Wel} {et~al.}(2008){van der Wel}, {Holden}, {Zirm}, {Franx},
  {Rettura}, {Illingworth}, \& {Ford}}]{vanderwel2008}
{van der Wel}, A., {Holden}, B.~P., {Zirm}, A.~W., {et~al.} 2008, \apj, 688, 48

\bibitem[{{van Dokkum} \& {Ellis}(2003)}]{vanDokkumEllis2003}
{van Dokkum}, P.~G. \& {Ellis}, R.~S. 2003, \apjl, 592, L53

\bibitem[{{van Dokkum} {et~al.}(2008){van Dokkum}, {Franx}, {Kriek}, {Holden},
  {Illingworth}, {Magee}, {Bouwens}, {Marchesini}, {Quadri}, {Rudnick},
  {Taylor}, \& {Toft}}]{vanDokkum2008}
{van Dokkum}, P.~G., {Franx}, M., {Kriek}, M., {et~al.} 2008, \apjl, 677, L5

\bibitem[{{van Dokkum} {et~al.}(2015){van Dokkum}, {Nelson}, {Franx}, {Oesch},
  {Momcheva}, {Brammer}, {F{\"o}rster Schreiber}, {Skelton}, {Whitaker}, {van
  der Wel}, {Bezanson}, {Fumagalli}, {Illingworth}, {Kriek}, {Leja}, \&
  {Wuyts}}]{vanDokkum2015}
{van Dokkum}, P.~G., {Nelson}, E.~J., {Franx}, M., {et~al.} 2015, \apj, 813, 23

\bibitem[{{van Dokkum} {et~al.}(2010){van Dokkum}, {Whitaker}, {Brammer},
  {Franx}, {Kriek}, {Labb{\'e}}, {Marchesini}, {Quadri}, {Bezanson},
  {Illingworth}, {Muzzin}, {Rudnick}, {Tal}, \& {Wake}}]{vanDokkum2010}
{van Dokkum}, P.~G., {Whitaker}, K.~E., {Brammer}, G., {et~al.} 2010, \apj,
  709, 1018

\bibitem[{{Vanzella} {et~al.}(2008){Vanzella}, {Cristiani}, {Dickinson},
  {Giavalisco}, {Kuntschner}, {Haase}, {Nonino}, {Rosati}, {Cesarsky},
  {Ferguson}, {Fosbury}, {Grazian}, {Moustakas}, {Rettura}, {Popesso},
  {Renzini}, {Stern}, \& {GOODS Team}}]{Vanzella2008}
{Vanzella}, E., {Cristiani}, S., {Dickinson}, M., {et~al.} 2008, \aap, 478, 83

\bibitem[{{Vanzella} {et~al.}(2005){Vanzella}, {Cristiani}, {Dickinson},
  {Kuntschner}, {Moustakas}, {Nonino}, {Rosati}, {Stern}, {Cesarsky}, {Ettori},
  {Ferguson}, {Fosbury}, {Giavalisco}, {Haase}, {Renzini}, {Rettura}, {Serra},
  \& {The Goods Team}}]{Vanzella2005}
{Vanzella}, E., {Cristiani}, S., {Dickinson}, M., {et~al.} 2005, \aap, 434, 53

\bibitem[{{Vanzella} {et~al.}(2006){Vanzella}, {Cristiani}, {Dickinson},
  {Kuntschner}, {Nonino}, {Rettura}, {Rosati}, {Vernet}, {Cesarsky},
  {Ferguson}, {Fosbury}, {Giavalisco}, {Grazian}, {Haase}, {Moustakas},
  {Popesso}, {Renzini}, {Stern}, \& {GOODS Team}}]{Vanzella2006}
{Vanzella}, E., {Cristiani}, S., {Dickinson}, M., {et~al.} 2006, \aap, 454, 423

\bibitem[{{Vazdekis}(1999)}]{Vazdekis1999}
{Vazdekis}, A. 1999, \apj, 513, 224

\bibitem[{{Vazdekis} {et~al.}(2010){Vazdekis}, {S{\'a}nchez-Bl{\'a}zquez},
  {Falc{\'o}n-Barroso}, {Cenarro}, {Beasley}, {Cardiel}, {Gorgas}, \&
  {Peletier}}]{Vazdekis2010}
{Vazdekis}, A., {S{\'a}nchez-Bl{\'a}zquez}, P., {Falc{\'o}n-Barroso}, J.,
  {et~al.} 2010, \mnras, 404, 1639

\bibitem[{{Wellons} {et~al.}(2016){Wellons}, {Torrey}, {Ma}, {Rodriguez-Gomez},
  {Pillepich}, {Nelson}, {Genel}, {Vogelsberger}, \& {Hernquist}}]{Wellons2016}
{Wellons}, S., {Torrey}, P., {Ma}, C.-P., {et~al.} 2016, \mnras, 456, 1030

\bibitem[{{Wellons} {et~al.}(2015){Wellons}, {Torrey}, {Ma}, {Rodriguez-Gomez},
  {Vogelsberger}, {Kriek}, {van Dokkum}, {Nelson}, {Genel}, {Pillepich},
  {Springel}, {Sijacki}, {Snyder}, {Nelson}, {Sales}, \&
  {Hernquist}}]{Wellons2015}
{Wellons}, S., {Torrey}, P., {Ma}, C.-P., {et~al.} 2015, \mnras, 449, 361

\bibitem[{{Whitaker} {et~al.}(2016){Whitaker}, {Bezanson}, {van Dokkum},
  {Franx}, {van der Wel}, {Brammer}, {Forster-Schreiber}, {Giavalisco},
  {Labbe}, {Momcheva}, {Nelson}, \& {Skelton}}]{Whitaker2016}
{Whitaker}, K.~E., {Bezanson}, R., {van Dokkum}, P.~G., {et~al.} 2016, ArXiv
  e-prints

\bibitem[{{Whitaker} {et~al.}(2012){Whitaker}, {Kriek}, {van Dokkum},
  {Bezanson}, {Brammer}, {Franx}, \& {Labb{\'e}}}]{Whitaker2012}
{Whitaker}, K.~E., {Kriek}, M., {van Dokkum}, P.~G., {et~al.} 2012, \apj, 745,
  179

\bibitem[{{Wilkes} {et~al.}(2013){Wilkes}, {Kuraszkiewicz}, {Haas}, {Barthel},
  {Leipski}, {Willner}, {Worrall}, {Birkinshaw}, {Antonucci}, {Ashby}, {Chini},
  {Fazio}, {Lawrence}, {Ogle}, \& {Schulz}}]{Wilkes2013}
{Wilkes}, B.~J., {Kuraszkiewicz}, J., {Haas}, M., {et~al.} 2013, \apj, 773, 15

\bibitem[{{Williams} {et~al.}(2014){Williams}, {Giavalisco}, {Cassata},
  {Tundo}, {Wiklind}, {Guo}, {Lee}, {Barro}, {Wuyts}, {Bell}, {Conselice},
  {Dekel}, {Faber}, {Ferguson}, {Grogin}, {Hathi}, {Huang}, {Kocevski},
  {Koekemoer}, {Koo}, {Ravindranath}, \& {Salimbeni}}]{Williams2014}
{Williams}, C.~C., {Giavalisco}, M., {Cassata}, P., {et~al.} 2014, \apj, 780, 1

\bibitem[{{Williams} {et~al.}(2015){Williams}, {Giavalisco}, {Lee}, {Tundo},
  {Mobasher}, {Nayyeri}, {Ferguson}, {Koekemoer}, {Trump}, {Cassata}, {Dekel},
  {Guo}, {Lee}, {Pentericci}, {Bell}, {Castellano}, {Finkelstein}, {Fontana},
  {Grazian}, {Grogin}, {Kocevski}, {Koo}, {Lucas}, {Ravindranath}, {Santini},
  {Vanzella}, \& {Weiner}}]{Williams2015}
{Williams}, C.~C., {Giavalisco}, M., {Lee}, B., {et~al.} 2015, \apj, 800, 21

\bibitem[{{Williams} {et~al.}(2010){Williams}, {Quadri}, {Franx}, {van Dokkum},
  {Toft}, {Kriek}, \& {Labb{\'e}}}]{Williams2010}
{Williams}, R.~J., {Quadri}, R.~F., {Franx}, M., {et~al.} 2010, \apj, 713, 738

\bibitem[{{Worthey} \& {Ottaviani}(1997)}]{WortheyOttaviani1997}
{Worthey}, G. \& {Ottaviani}, D.~L. 1997, \apjs, 111, 377

\bibitem[{{Xue} {et~al.}(2011){Xue}, {Luo}, {Brandt}, {Bauer}, {Lehmer},
  {Broos}, {Schneider}, {Alexander}, {Brusa}, {Comastri}, {Fabian}, {Gilli},
  {Hasinger}, {Hornschemeier}, {Koekemoer}, {Liu}, {Mainieri}, {Paolillo},
  {Rafferty}, {Rosati}, {Shemmer}, {Silverman}, {Smail}, {Tozzi}, \&
  {Vignali}}]{Xue2011}
{Xue}, Y.~Q., {Luo}, B., {Brandt}, W.~N., {et~al.} 2011, \apjs, 195, 10

\bibitem[{{Yan} \& {Blanton}(2012)}]{YanBlanton2012}
{Yan}, R. \& {Blanton}, M.~R. 2012, \apj, 747, 61

\bibitem[{{Yano} {et~al.}(2016){Yano}, {Kriek}, {van der Wel}, \&
  {Whitaker}}]{Yano2016}
{Yano}, M., {Kriek}, M., {van der Wel}, A., \& {Whitaker}, K.~E. 2016, \apjl,
  817, L21

\bibitem[{{Zirm} {et~al.}(2007){Zirm}, {van der Wel}, {Franx}, {Labb{\'e}},
  {Trujillo}, {van Dokkum}, {Toft}, {Daddi}, {Rudnick}, {Rix},
  {R{\"o}ttgering}, \& {van der Werf}}]{Zirm2007}
{Zirm}, A.~W., {van der Wel}, A., {Franx}, M., {et~al.} 2007, \apj, 656, 66

\bibitem[{{Zolotov} {et~al.}(2015){Zolotov}, {Dekel}, {Mandelker}, {Tweed},
  {Inoue}, {DeGraf}, {Ceverino}, {Primack}, {Barro}, \& {Faber}}]{Zolotov2015}
{Zolotov}, A., {Dekel}, A., {Mandelker}, N., {et~al.} 2015, \mnras, 450, 2327

\end{thebibliography}

\end{document}